\documentclass[usenatbib]{mnras}
\hypersetup{draft}

\usepackage{newtxtext,newtxmath}
\usepackage[T1]{fontenc}
\usepackage[utf8]{inputenc}
\usepackage{ae,aecompl}

\usepackage[usenames, dvipsnames]{color}
\usepackage{graphicx}	\usepackage{amsmath}	\usepackage{amssymb}	\usepackage{pifont}
\usepackage{array}
\usepackage{hyperref}
\usepackage{cleveref}

\newcommand{\km}{\ensuremath{\, \mathrm{km}}}
\newcommand{\cm}{\ensuremath{\, \mathrm{cm}}}
\newcommand{\pc}{\ensuremath{\, \mathrm{pc}}}

\newcommand{\Mpc}{\ensuremath{\, \mathrm{Mpc}}}

\newcommand{\Msun}{\ensuremath{\, \mathrm{M}_{\odot}}}
\newcommand{\Rsun}{\ensuremath{\, \mathrm{R}_{\odot}}}
\newcommand{\s}{\ensuremath{\, \mathrm{s}}}
\newcommand{\yr}{\ensuremath{\, \mathrm{yr}}}
\newcommand{\kyr}{\ensuremath{\, \mathrm{kyr}}}
\newcommand{\Myr}{\ensuremath{\, \mathrm{Myr}}}
\newcommand{\Gyr}{\ensuremath{\, \mathrm{Gyr}}}
\newcommand{\G}{\ensuremath{\, \mathrm{G}}}

\newcommand{\K}{\ensuremath{\, \mathrm{K}}}
\newcommand{\kms}{\ensuremath{\km \s^{-1}}}
\newcommand{\amu}{\ensuremath{\, \mathrm{amu}}}
\newcommand{\erg}{\ensuremath{\, \mathrm{erg}}}
\newcommand{\keV}{\ensuremath{\, \mathrm{keV}}}

\newcommand{\mo}{\ensuremath{^{-1}}}
\newcommand{\ie}{\emph{i.e.}\, }
\newcommand{\eg}{\emph{e.g.}\, }
\newcommand{\ssi}{\ensuremath{\Leftrightarrow}}
\renewcommand{\d}{\ensuremath{\mathrm{\, d}}}
\renewcommand{\log}{\ensuremath{\mathrm{log}_{10}}}

\newcommand{\Mbh}{\ensuremath{M_\bullet}}
\newcommand{\mstar}{\ensuremath{m_\star}}
\newcommand{\rstar}{\ensuremath{r_\star}}
\newcommand{\rc}{\ensuremath{r_c}}
\newcommand{\Ramses}{\textsc{Ramses}}
\newcommand{\vv}[1][args]{\mathbf}

\title[TDEs under the grid]{Tidal disruption events in the first billion years of a galaxy}

\author[H. Pfister et al.]{Hugo Pfister,$^{1,2}$\thanks{Sophie and Tycho Brahe Fellow; hugo.pfister@nbi.ku.dk}
Jane Dai,$^{2,1}$ Marta Volonteri,$^{3}$ Katie Auchettl,$^{5,6,1,4}$\newauthor Maxime Trebitsch$^{7,8}$ and Enrico Ramirez-Ruiz$^{1,4}$
\\
$^{1}$DARK, Niels Bohr Institute, University of Copenhagen, Blegdamsvej 17, DK-2100 Copenhagen, Denmark\\
$^{2}$Department of Physics, The University of Hong Kong, Pokfulam Road, Hong Kong, China\\
$^{3}$Institut d'Astrophysique de Paris, Sorbonne Universit\'{e}, CNRS, UMR7095, 98bis boulevard Arago, F-75014, Paris, France\\
$^{4}$Department of Astronomy and Astrophysics, University of California, Santa Cruz, CA 95064, USA\\
$^{5}$School of Physics, The University of Melbourne, Parkville, VIC 3010, Australia\\
$^{6}$ARC Centre of Excellence for All Sky Astrophysics in 3 Dimensions (ASTRO 3D)\\
$^{7}$Max-Planck-Institut f{\"u}r Astronomie, K{\"o}nigstuhl 17, 69117 Heidelberg, Germany\\
$^{8}$Zentrum f{\"u}r Astronomie der Universit{\"a}t Heidelberg, Institut f{\"u}r Theoretische Astrophysik, Albert-Ueberle-Str. 2, 69120 Heidelberg, Germany
}

\date{Accepted XXX. Received YYY; in original form ZZZ}

\pubyear{2019}

\begin{document}
\label{firstpage}
\pagerange{\pageref{firstpage}--\pageref{lastpage}}
\maketitle

\begin{abstract}
Accretion of stars on massive black holes (MBHs) can feed MBHs and generate tidal disruption events (TDEs). We introduce a new physically motivated model to self-consistently treat TDEs in cosmological simulations, and apply it to the assembly of a galaxy with final mass $3\times 10^{10}\Msun$ at $z=6$. This galaxy exhibits a TDE rate of $\sim 10^{-5}\yr\mo$, consistent with local observations but already in place when the Universe was one billion year old. A fraction of the disrupted stars participate in the growth of MBHs, dominating it until the MBH reaches mass $\sim 5 \times 10^5 \Msun$, but their contribution then becomes negligible compared to gas. TDEs could be a viable mechanism to grow light MBH seeds, but fewer TDEs are expected when the MBH becomes sufficiently massive to reach the luminosity of, and be detected as, an active galactic nucleus. Galaxy mergers bring multiple MBHs in the galaxy, resulting in an enhancement of the global TDE rate in the galaxy by $\sim 1$ order of magnitude during $100\Myr$ around mergers. This enhancement is not on the central MBH, but caused by the presence of MBHs in the infalling galaxies.  This is the first self consistent study of TDEs in a cosmological environment and highlights that accretion of stars and TDEs are a natural process occurring in a Milky~Way-mass galaxy at early cosmic times.
\end{abstract}


\section{Introduction}

Gas falling onto black holes, increasing their mass and releasing gravitational energy, can explain the growth of massive black holes (MBHs) with masses $\gtrsim 10^6\Msun$ in the center of most massive galaxies \citep{Kormendy_13}. However, the energy released from gas falling onto MBHs as well as from nearby supernovae (also known as ``feedback'') can heat and eject gas, preventing MBH growth in low-mass galaxies \citep[\eg ][]{Dubois_15, Habouzit_17,Trebitsch_18}. As a result, it is challenging to explain, with gas accretion only, the presence of MBHs with masses $\gtrsim 10^9\Msun$ at $z>6$ when the Universe was only 1~Gyr \citep{Tenneti_17,Banados_18}.

However, the vicinity of MBHs is not only composed of gas, but also of stars \citep{Schodel_18} that can also be accreted by MBHs and increase their mass. Furthermore, contrary to gas, stars are not subject to feedback, therefore they could provide a continuous source of material for MBHs to accrete and grow \citep{Alexander_17}. In order to know the contribution of stars to the growth of MBHs, one needs to know the rate at which stars get close enough to a MBH to be swallowed, whole or in part, and increase its mass. For a Solar-like star and MBHs with mass $\lesssim 10^8\Msun$, stars are not swallowed whole but tidally disrupted, producing a unique signature known as a tidal disruption event \citep[TDE; ][]{Lacy_82,Rees_88}. This allows us to observationally measure the TDE rate, thus providing an estimate of the rate at which stars get close enough to MBHs to increase their masses.

With a handful of observed TDEs, for central massive MBHs in quiescent galaxies at $z=0$, the typical rate is $(0.1-1.7)\times 10^{-4}\yr^{-1}$ \citep{Donley_02, Gezari_08, VanVelzen_14, Holoien_16, Blagorodnova_17, Auchettl_18, VanVelzen_18}. This average rate can be well understood theoretically with the analytical loss-cone theory \citep{Lightman_77, Magorrian_99, Wang_04}: a MBH with mass $M_\bullet$ is embedded in a stellar density profile and fed by stars scattered toward its direction through 2-body interactions. This formalism also predicts that there is a negative correlation between the TDE rate and the mass of the MBH \citep[the TDE rate scales as $M^{-\epsilon}_\bullet$ with $0< \epsilon < 0.5$, ][]{Wang_04, Stone_16a, Pfister_20b}, which is confirmed by observations for MBHs with a mass $\gtrsim 10^6\Msun$ \citep{VanVelzen_18}.

This justifies the use of the the loss-cone theory to estimate the growth of MBHs through TDEs, and \cite{Alexander_17} showed that MBHs can reach masses above $\sim 3 \times 10^5\Msun$ regardless the initial MBH seed mass and redshift formation only through accretion of stars. While this suggests that TDEs are efficient in growing light MBH seeds, this result was derived under idealized assumptions: MBHs are embedded in a singular isothermal sphere \citep{BT_87} with an inner Bahcall-Wolf cusp \citep{Bahcall_76}, the $\Mbh-\sigma$ relation \citep{Ferrarese_02, Kormendy_13} is always verified, MBHs are fixed in the center of the stellar distribution and the only relevant process is 2-body interactions. To relax these idealized assumptions, N-body simulations in which stars getting close enough to the MBH can be directly counted have been carried out \citep{Baumgardt_04a, Baumgardt_04b, Brockamp_11, Zhong_14}. \cite{Baumgardt_04a} and \cite{Brockamp_11} find that MBHs with masses $\gtrsim 10^3\Msun$ can only double their mass within a Hubble time. To summarize, the contribution of TDEs to the growth of MBHs is still uncertain.

Furthermore, these previous studies assume that MBHs and their host galaxies are isolated, while most galaxies undergo several mergers during their life \citep[\eg ][]{Fakhouri_10}. These mergers drastically affect galaxies, triggering star formation and substantially changing the stellar density profile near the central MBHs \citep[\eg ][]{VanWassenhove_14, Capelo_15}. As a consequence, it is natural to believe that galaxy mergers affect the TDE rate. In addition, this enhancement of the TDE rate during galaxy mergers is somewhat motivated by observations, as E+A galaxies which are post-mergers galaxies, are found to have an enhanced TDE rate of $10^{-3}\yr\mo$ \citep{French_16, Stone_16b}. To test this, N-body simulations of galaxy mergers have been performed \citep[\eg][]{LiShuo_17, Sakurai_18}, and they indeed find that mergers enhance the TDE rate. There are two reasons to this, and both of them result in an enhancement of the loss cone feeding: \textit{(i)} the stellar distribution is overall more triaxal due to the merger; and \textit{(ii)} when the MBHs get close to each other, stars bound to one MBH see their dynamics greatly perturbed by the companion MBH.

However, neither the loss-cone formalism nor N-body simulations include gas, which can cool and turn into new stars that can then be disrupted. Therefore these frameworks cannot provide a fully consistent picture. To partially overcome this issue, \cite{Pfister_19b} adopted a trade-off between the ability to estimate exactly the TDE rate by resolving stars getting close enough to MBHs and including the physics of galaxies (star formation, supernovae etc...): they used an isolated hydrodynamical simulation of a galaxy merger starting from idealized initial conditions, and post-processed the TDE rate applying loss-cone theory \citep{Vasiliev_17, Vasiliev_18} onto the self-consistently evolving density profiles. They found that, indeed, during mergers, nuclear starbursts around MBHs enhance the central stellar density, naturally resulting in an enhancement of the TDE rate \citep{Stone_16b}. 

Furthermore, as the TDE rate results from a combination of the properties of the MBH and its surrounding stellar density profile, it is natural that it varies from galaxy to galaxy \citep{French_20b}. We mentioned the enhancement in post-mergers E+A galaxies, but ultra-luminous infrared galaxies could have a TDE rate as high as $10\mo\yr\mo$ \citep{Tadhunter_17, Kool_20}, and high redshift galaxies which are  more star forming and compact \citep[e.g.,][]{Madau_14, Allen_17} as well active galactic nuclei (AGNs) could also exhibit a different TDE rate.

The current status of the field hints to a diversity of TDE rates based on galaxy properties, environment and cosmic epoch. The role and importance of stellar accretion on MBH growth and the evolution of the TDE rate must be investigated in a fully cosmological context, in which galaxies grow over time by accretion of cosmic filaments, where galaxy mergers are numerous, especially at early cosmic times, and galaxies are more ``messy'' than in an ideal set-up (compare Fig.~3 of \cite{Capelo_15} with Fig.~\ref{fig:snapshots}).

In this paper, we introduce in \S\ref{sec:TidalDisruptionEventsUnderTheGrid} a new subgrid model to self-consistently take into account TDEs in cosmological simulations, and we apply it to the assembly of a galaxy with final mass ${3\times 10^{10}\Msun}$ at $z=6$ described in \S\ref{sec:NumericalSetUp}. This allows us not only to study the contribution of TDEs to the growth of MBHs, but also the evolution of the TDE rate during mergers and AGN phases, as this galaxy suffers several mergers and sometime has an AGN. We discuss our results in \S\ref{sec:Results} and conclude in \S\ref{sec:Conclusions}.

\section{Tidal disruption events under the grid}
\label{sec:TidalDisruptionEventsUnderTheGrid}

We present how we estimate the TDE rate given the properties around MBHs. We first recall analytical estimates in \S\ref{sec:theory}, we then detail in \S \ref{sec:Implementation} the implementation in \Ramses\, \citep{Teyssier_02} and finish by the caveats of our model in \S\ref{sec:Caveats}.

\subsection{Theory}
\label{sec:theory}
It is customary to express TDEs as sourced by two different regions \citep{Syer_99,Merrit_book}: the empty loss cone \citep{Wang_04}, close to the MBH ($r<r_c$, $r_c$ is defined in the following paragraph), where the diffusion timescale $T_r$  is longer than the radial period; and from the full loss cone \citep{Pfister_19b}, farther away ($r>r_c$),  where the diffusion timescale is shorter than the radial period.

We assume a MBH with a mass \Mbh, embedded in a stellar density and stellar velocity dispersion profiles $\rho$ and $\sigma$, all stars  having a mass $\mstar$ and radius  $\rstar$. In this situation, $r_c$ is the radius at which the contributions of the full and empty loss cone to the flux of stars match, meaning that $r_c$ is solution to \citep{Pfister_19b}:
\begin{eqnarray}
&&\frac{\G \rho(\rc) \rc ^3}{\sigma^2(\rc)} = \rstar q^{4/3} \\ 
&\ssi& \frac{ \rho(\rc) \rc ^4}{M(\rc) + \Mbh} = \rstar q^{4/3} \label{eq:rcrit}
\, ,
\end{eqnarray}
where $q=\Mbh/ \mstar$; and we have assumed the velocity dispersion to be $\sigma(r)^2 \sim \G (M(r)+\Mbh) / r$, where $M(r)$ is the enclosed stellar mass within $r$.

Following \cite{Wang_04}, we estimate the TDE rate coming from the empty loss cone as:
\begin{eqnarray}
\Gamma_{\rm empty} &=& \frac{M(r_c)}{\mstar T_r(r_c)} \, ,
\end{eqnarray}
where $T_r$ is the \citep{Spitzer_58}:
\begin{eqnarray}
T_r(r) = \frac{\sqrt{2} \sigma^3(r)}{\pi \G^2 \mstar \rho(r) \ln(0.4 \Mbh / \mstar )} \,.
\end{eqnarray}

Following \cite{Pfister_19b}, we estimate the TDE rate coming from the full loss cone as:
\begin{eqnarray}
\Gamma_{\rm full} &=& 2 \pi \G q^{4/3} \rstar \frac{\rho(\rc)}{\sigma(\rc)} \, .
\end{eqnarray}

The total TDE rate, $\Gamma$, can be expressed as the sum of the two:
\begin{eqnarray}
\Gamma = \Gamma_{\rm empty} + \Gamma_{\rm full} \label{eq:GammaTot} \, .
\end{eqnarray}

To get a step further, we assume the stellar density profile to be a power law, with logarithmic slope $-3 < \gamma \leq 0$:
\begin{eqnarray} 
\rho(r)=\rho_0\frac{3-\gamma}{3} \left( \frac{r}{r_0}\right)^{-\gamma} \label{eq:density} \\
M(r) = \frac{4}{3}\pi r_0^3 \rho_0  \left( \frac{r}{r_0}\right)^{3-\gamma} \, ,
\end{eqnarray}
where $\rho_0$ corresponds to the mean density within $r_0$. In this situation, we can rewrite Eq.~\eqref{eq:rcrit} as:
\begin{eqnarray}
1+\tilde{\rho}_0 \tilde{r}^{3-\gamma}_c=2 \tilde{\rho}_0 \tilde{r}^{4-\gamma}_c \, \label{eq:rcrit_polynomial}
\end{eqnarray}
where we introduce $\tilde{\rho}_0 = \rho_0 / \rho_u $ and $\tilde{r}_c  = r_c / r_u$, with:

\begin{eqnarray}
\rho_{u} &=& \frac{\mstar}{4/3\pi \rstar^3} \left( \frac{\rstar}{r_0}\right)^\gamma \left( \frac{3-\gamma}{8\pi}\right)^{3-\gamma} q^{(4\gamma-9)/3} \\
r_u &=& \frac{8 \pi}{3-\gamma} q^{4/3} \rstar   \, .
\end{eqnarray}

Unfortunately, even for this simple density profile, in general, no explicit expression of \rc\ can be written. However, in some limiting regimes we have:

\begin{eqnarray}
	r_c \sim \left\{
	\begin{aligned}
	&r_u \left( \frac{2 \rho_0}{\rho_{u}} \right)^{1/(-4+\gamma)}	&\textrm{ if }&  \rho_0 \ll \rho_{u} \\
    & r_u &\textrm{ if }& \rho_0 = \rho_{u}	\\
	&\frac{r_u}{2} &\textrm{ if }& \rho_0 \gg \rho_{u}	\\
	\end{aligned}
	\right . 
	\, .
\end{eqnarray}

allowing us to estimate the TDE rate.

\subsection{Implementation}
\label{sec:Implementation}

Here we detail how we go from the theoretical analysis derived in \S\ref{sec:theory}, to the actual implementation in \Ramses.

At each timestep of the simulation, between $t$ and $t+\Delta t$, we estimate the mean stellar density, $\rho_{0,\mathcal{S}_k}$ ($k$ is 4 or 8), in the sphere $\mathcal{S}_k$, of radius $k\Delta x$ centered on the MBH ($\Delta x$ being the minimum cell size in the simulation) as:
\begin{eqnarray}
\rho_{0,\mathcal{S}_k} &=& \frac{1}{4/3 \pi (k \Delta x)^3} \sum_{i\in\mathcal{S}_k}m_i \\
     &=& \frac{1}{4/3 \pi (k \Delta x)^3} M_k\, ,
\end{eqnarray}
where $m_i$ is the mass of the stellar particle $i$; and $M_k$ is the total stellar mass enclosed within $k \Delta x$ around the MBH. If we assume the density around the MBH to be expressed as given in Eq.~\eqref{eq:density}, we obtain:
\begin{eqnarray}
\rho_0 &=& \rho_{0,\mathcal{S}_4} \\
r_0 &=& 4 \Delta x \\
\gamma &=& 3-\ln_2 \left( \frac{M_8}{M_4} \right) \label{eq:gammaslope} \, .
\end{eqnarray}

The mass of the MBH, $\Mbh$, is measured directly in the simulation, and we assume stars to be all solar--like, that is $\rstar=\Rsun$ and $\mstar=\Msun$. Finally, we estimate $r_c$ as:
\begin{eqnarray}
r_c &=& r_u \left[ \left( \frac{2 \rho_0}{\rho_{u}} \right)^{1/(-4+\gamma)} \frac{-\tanh\left( \log\left( \frac{\rho_0}{\rho_{u}} \right) \right) +1 }{2} + \right. \nonumber \\
&&  \frac{1}{2} \frac{\tanh\left( \log\left( \frac{\rho_0}{\rho_{u}} \right) \right) +1 }{2} + \label{eq:rcrit_fit}\\ 
&&\left. \left( \frac{3}{4} - 2^{-1+1/(-4+\gamma)} \right) \exp\left( -  \log^2\left( \frac{\rho_0}{\rho_{u}} \right)  \right) \right] \, , \nonumber
\end{eqnarray}
which approximates the true value of $r_c$, solution to Eq.~\eqref{eq:rcrit}, within less than 30\% error (see Appendix~\ref{sec:AnEstimateOfTheCriticalRadius}).

With all this, we can estimate the TDE rate onto the MBH, $\Gamma$ (Eq.~\eqref{eq:GammaTot}), as shown in \S\ref{sec:theory}.

A mass $\Gamma \mstar  \Delta t$  is then removed from surrounding stars within $4 \Delta x$ and the three following steps are done:
\begin{enumerate}
\item A mass $\dot{M}_{\bullet, \textrm{star}} \Delta t=f_a \Gamma \mstar (1-\epsilon_r) \Delta t $ is added to the MBH, where $f_a=0.5$ is the fraction of mass which falls onto the MBH\footnote{Note that in the paper we clearly make the difference between the TDE rate in $\yr^{-1}$ corresponding to the number of stars being disrupted, and the ``stellar accretion rate'' (stars are not accreted \textit{per se}, gas falling back from the disrupted stars is) in $\Msun \yr^{-1}$ corresponding to the total mass of disrupted stars falling onto the MBH. This difference is mainly ``syntactic'' as we assumed that all stars are solar like and $f_a=0.5$, therefore the stellar accretion rate and the TDE rate differ by a factor of two in their respective units.}; and  $\epsilon_r$ is the radiative efficiency, which depends on the spin of the MBH (4\% for a non rotating MBH and up to 42\% for a highly spinning MBH; in this paper we use fixed value of $\epsilon_r=10\%$, see \S\ref{sec:BHImplementation}).
\item A mass $\dot{M}_{d} \Delta t=(1-f_a) \Gamma \mstar \Delta t $ does not fall onto the MBH and returns into cells containing disrupted stars as gas (see Eq.~\eqref{eq:gasTDE}).
\item An energy $f_a \Gamma \mstar \epsilon_r \Delta t c^2$ is emitted by the MBH. At the moment,  we consider that the feedback is similar for accreted stars than for accreted gas: either thermal or kinetic depending on the Eddington ratio (see \S\ref{sec:AGNfeedback} for details on the implementation in \Ramses). Assuming a similar expression for the feedback following gas or stellar accretion is not absurd, indeed, once the accretion disk is formed following the disruption of a star, whether the material was originated from a star or a gaseous clump should not change the behavior. Note that this should be a upper limit of the radiative feedback, since the radiative efficiency likely has a smaller value than the thin disk one, and a fraction of the bound stellar debris can become outflows.
\end{enumerate}

In addition, to conserve total momentum, we update the velocity of the MBH, gas and stars accordingly. In the end, we have:
\begin{eqnarray}
&\textrm{stars} &\left\{
\begin{aligned}
m_i (t+\Delta t) &= m_i (t) -  \Gamma \mstar  \Delta t f_i \\
\vv{v_i}(t+\Delta t)&= \vv{v_i}(t)
\end{aligned} \right. \\
&\textrm{gas}& \left\{
\begin{aligned}
\rho_{\mathrm{g}, i}(t+\Delta t) &=  \rho_{\mathrm{g}, i}(t) + \frac{\dot{M}_{d}  \Delta t f_i }{\Delta x^3}\\
\vv{u_i}(t+\Delta t) &=  \frac{ \rho_{\mathrm{g}, i}(t)   \Delta x^3 \vv{u_i}(t) +\dot{M}_{d}  \Delta t f_i \vv{v_i}(t) }{\rho_{\mathrm{g}, i}(t) \Delta x^3 + \dot{M}_{d}  \Delta t f_i} \\
e_i(t+\Delta t)&= e_i(t) + \frac{1}{2} \frac{\dot{M}_{d} \Delta t f_i }{\Delta x^3} \vv{v_i}(t)^2
\end{aligned} \right. \label{eq:gasTDE} \\
&\textrm{MBH}& \left\{
\begin{aligned}
\vv{v}_\bullet (t+\Delta t)&= \frac{\vv{v}_\bullet(t) \Mbh(t) + \langle \vv{v}_\star \rangle \frac{\dot{M}_{\bullet, \textrm{star}}  \Delta t}{M_4}}{ \Mbh(t) + \dot{M}_{\bullet, \textrm{star}}  \Delta t}  \\
\Mbh(t+\Delta t) &= \Mbh(t) + \dot{M}_{\bullet, \textrm{star}}  \Delta t
\end{aligned} \right. \, ,
\end{eqnarray}
where $ \langle \vv{v}_\star \rangle $ is the mass-weighted velocity of stars, with velocities $\vv{v_i}$, within $4\Delta x$ from the MBH; $\vv{v}_\bullet$ is the velocity of the MBH; $ f_i = m_i(t)/M_4 $ is the contribution of the stellar particle $i$ to the TDE rate ($M_4=M_k$ for $k=4$ is the enclosed stellar mass within $4\Delta x$); and $\rho_{\mathrm{g}, i}$, $\vv{u_i}$ and $e_i$ are respectively the density, velocity and total energy density of the cell containing the stellar particle $i$.

\subsection{Caveats}
\label{sec:Caveats}

We discuss here a few numerical and physical caveats of the implementation:
\begin{itemize}
\item If the MBH is off-center from its host galaxy, and is therefore not in a spherical density profile, Eq.~\eqref{eq:gammaslope} could give negative $\gamma$. When $\gamma < 0$, we set $\Gamma = 0$.
\item If the available mass of stars ($M_4$) is lower than the disrupted mass $\Gamma \Delta t$, then there are not enough stars. In this situation, we set $\Gamma = M_4 / \Delta t$ and remove all available stars (note that in practice this did not happen in our simulations).
\item Even if the density profile is spherical around the MBH, it is possible that it does not follow a simple power law. Our ``bet'' is that, if the resolution of the simulation is high enough, then the estimate of the inner slope $\gamma$ is enough for an estimate of the TDE rate.  In practice as our simulation reaches a resolution $\Delta x \sim 7 \pc$ (see Table.~\ref{table:sims}), this translates into assuming a constant slope within $\sim 60\pc$ for our estimate of the TDE rate. Note that observed galaxies at $z \ll 1$ are usually well fitted with fixed inner slope within $\sim 100 \pc$ \citep[\eg][]{Lauer_07}, and there seem to be a correlation between density at these scales and the TDE rate \citep{French_20a}. However, this excludes the presence of a nuclear star cluster around MBHs \citep{Pechetti_19, SanchezJanssen_19} which could enhance the TDE rate by orders of magnitude \citep{Pfister_20b}.
\item It is currently not known what is the fraction of disrupted material which falls back onto the MBH ($f_a$), nor how long it takes. If a star comes with a highly parabolic orbit, \ie with total energy ``close to zero'', then we expect half of the debris to remain bound and half to be unbound. We assume  here that all bound debris immediately falls back onto the MBH ($f_a=0.5$).
\item  \cite{Pfister_19b} and \cite{Wang_04} only give an approximate TDE rate in the full and empty loss cone regime. More detailed analytical framework exist \citep{Stone_16a,Vasiliev_17}, but it would be numerically inefficient (it involves computing numerous ``integrals'') and meaningless (we assume a spherical density profile and all stars a solar like which are ``larger'' approximations than the full/empty loss cone) to use them.
\item Assuming that all stars are all solar--like is clearly simplistic, however, \cite{Stone_16a} have shown that using a stellar mass distribution function varies the TDE rate by only $\sim 2$ with respect to the monochromatic Solar population we consider.
\item Although stellar accretion can be super--Eddington, we still use the feedback thermal mode (see \S\ref{sec:BHImplementation}) from \cite{Dubois_12}. This is somewhat inconsistent with high resolution simulations close to the vicinity of the MBH \citep{Sadowski_16,Dai_18} which find that at high accretion rate, the feedback is more likely to be mechanical and possibly jetted if the conditions are optimal. We leave this development of super--Eddington accretion to a future study.
\end{itemize}

\section{Numerical  set-up}
\label{sec:NumericalSetUp}

In order to study the evolution of TDE rate in a galaxy evolving in a realistic context, we run a cosmological zoom on a halo whose properties are described in \S\ref{sec:ICs}. The simulation is performed with the publicly available adaptive mesh refinement (AMR) code \Ramses\, \citep{Teyssier_02}.

\Ramses\, follows the evolution of the gas using the second-order MUSCL-Hancock scheme for the Euler equations; and the approximate Harten-Lax-Van Leer Contact Riemann solver, with a MinMod total variation diminishing scheme to reconstruct the interpolated variables from their cell-centered values, is used to compute the unsplit Godunov fluxes at cell interfaces \citep{Toro_97}. An equation of state of perfect gas composed of monoatomic particles with adiabatic index $5/3$ is assumed to close the full set of fluid equations. The Courant factor is set to 0.8 to define the timestep. 

Collisionless particles (dark matter, stars and MBHs) are evolved using a particle-mesh solver with a cloud-in-cell (CIC) interpolation. The size of the CIC is that of the local cell for MBHs and stars. As dark matter (DM) particles are larger in mass, we smooth their distribution to reduce their contribution to shot noise, and they can only project their mass on the grid down to a minimum cell size of $\Delta x_\textrm{DM}$, corresponding to the highest level unlocked when running the DM only simulation with the same mass resolution.

The AMR grid is refined using a quasi-Lagrangian criterion: a cell is refined if $M_\textrm{DM}^\textrm{cell}+ (\Omega_m / \Omega_b -1) M_b^\textrm{cell} \geq 8 \times m^\textrm{part}_\textrm{DM} $, where $M_\textrm{DM}$ and $M_b^\textrm{cell} $ are respectively the mass of dark matter and baryons in the cell;  $\Omega_m$ and $\Omega_b$ are the total matter and baryon density and $m^\textrm{part}_\textrm{DM}$ is the mass of high-resolution dark matter particles. The minimum cell size, $\Delta x$, is kept roughly constant in proper physical size with redshift: an additional level of refinement is added every time the expansion factor $a_\textrm{exp}$ increases by a factor of two, such that the maximum level, $l_\textrm{max}$, is reached at $a_\textrm{exp} = 0.8$. For simplicity, we further assume that $\Delta x=L_\textrm{box}/2^{l_\textrm{max}}$, where $L_\textrm{box}$ is the size of the box at $z=0$.

The subgrid physics is described below in \S\ref{sec:GalImplementation} and \S\ref{sec:BHImplementation}, a summary of main quantities of the simulation can be found in Table~\ref{table:sims}.

\begin{table}
 \begin{tabular}{|c|l|r|}
 \hline
 \textbf{Name} & \textbf{Value} & \textbf{Comments} \\
 \hline
 $L_\mathrm{box}$ & 59 Mpc  & Size of the box at $z=0$ \\
 $M_\mathrm{vir}$ & $3\times 10^{11}$ & at $z=5.7$ \\
 $l_\mathrm{max}$ & 23 & Maximum level of refinement of the AMR grid \\
 $\Delta x$ & 7 pc & Best spatial resolution \\
 $\Delta x_\mathrm{DM}$ & 450\pc & Spatial resolution of dark matter \\
 $m^\mathrm{part}_\mathrm{DM}$ & $10^5 \Msun$  & Mass of high resolution dark matter particles \\
 $m^\mathrm{part}_\star$ & $6\times 10^3 \Msun$  & Mass of stellar particles \\
 $M_{\bullet, \mathrm{seed}}$ & $10^5 \Msun$  & Seed mass of MBHs \\
\hline 
 \end{tabular}
 \caption{Simulation parameters}
 \label{table:sims}
\end{table}

\subsection{Initial conditions}
\label{sec:ICs}

The initial conditions are produced with  \textsc{Music} \citep{MUSIC} and are the same as in \cite{Trebitsch_19}. We assume a $\Lambda$CDM cosmology with total matter density $\Omega_m=0.3089$, baryon density $\Omega_b = 0.0486$, dark energy density $\Omega_\Lambda = 0.6911$, amplitude of the matter power spectrum $\sigma_8 = 0.8159$, $n_s = 0.9667$ spectral index and Hubble constant $H_0 = 67.74 \km \s^{-1}  \Mpc^{-1}$ consistent with the \emph{Planck} data \citep{Planck_15}.

Low resolution dark matter particles with mass $m^\textrm{coarse}_\textrm{DM} = 4\times10^8\Msun$ are placed onto the box with an effective resolution of $256^3$ elements. Additional high-resolution dark matter particles, with an effective resolution of $4096^3$ elements corresponding to a mass $m^\textrm{part}_\textrm{DM}=10^5\Msun$, are placed around a halo of mass $M_\textrm{vir} = 3\times 10^{11} \Msun$ at $z = 5.7$.

\subsection{Physics of galaxies}
\label{sec:GalImplementation}

\subsubsection{Cooling and heating}

Gas is allowed to cool by hydrogen and helium with a contribution from metals using cooling curves from \cite{Sutherland_93} for temperatures above $10^4 \K$. For gas below $10^4 \K$ and down to our minimum temperature of $10 \K$, we use the fitting functions of~\cite{Rosen_95}.

The effect of reionization is modelled with a uniform heating from the UVB background from \cite{Haardt_96} below $z=8.5$. In addition, to take into account that the center of dense regions can be shielded by neutral hydrogen, the UV photo-heating is reduced by $\exp(-\rho_\mathrm{g}/\rho_\mathrm{shield})$, where $\rho_\mathrm{g}$ is the gas density of the cell and $\rho_\mathrm{shield}=0.01 \amu \cm^{-3}$.

\subsubsection{Star formation}

During each timestep $\Delta t$, in leaf cells with gas density $\rho_\textrm{g}~>~1\amu\cm^{-3}$,  $N$ stellar particles with mass $m^\textrm{part}_\star=~6\times~10^3\Msun$ are drawn from a Poisson distribution with parameter $\lambda = M_\textrm{SF}/m^\textrm{part}_\star$, where $M_\textrm{SF}$ is the mass of newly formed stars \citep{Rasera_06}. $M_\textrm{SF}$ is computed so that the star formation rate follows a Kennicutt--Schmidt Law \citep{Schmidt_59,Kennicutt_98}, that is $M_\textrm{SF}=\epsilon \rho_\textrm{g} \Delta x^3 \Delta t / t_\textrm{ff}$, where $\epsilon$ is the star formation efficiency and $t_\textrm{ff}=\sqrt{3\pi/(32\G \rho_\textrm{g})}$ is the free fall time. 

$\epsilon$ depends on the local properties of gas and is estimated using the multi-ff PN model from \cite{Federrath_12}.

\subsubsection{Stellar feedback}

21\% of the mass of each stellar particles\footnote{This corresponds to the mass fraction of stars more massive than $8\Msun$ assuming a Kroupa initial mass function \citep{Kroupa_01} with stars having a mass in between 0.08 and $100\Msun$.} is re-emitted in the medium in supernovae 5 Myr after their formation, releasing a (kinetic) energy of $2\times10^{49}\erg \Msun^{-1}$. The amount of momentum depositted depends on the local density and metallicity of each neighbouring cell, and depends on the stages of the Sedov-Taylor blast wave \citep[see ][]{Kimm_14}. In addition, modifications from  \cite{Kimm_17}, using the results of \cite{Geen_15}, to take into account pre-heating of the interstellar medium by radiation before the supernovae explosion, are used.

\begin{figure*}
\includegraphics[width=0.49\textwidth]{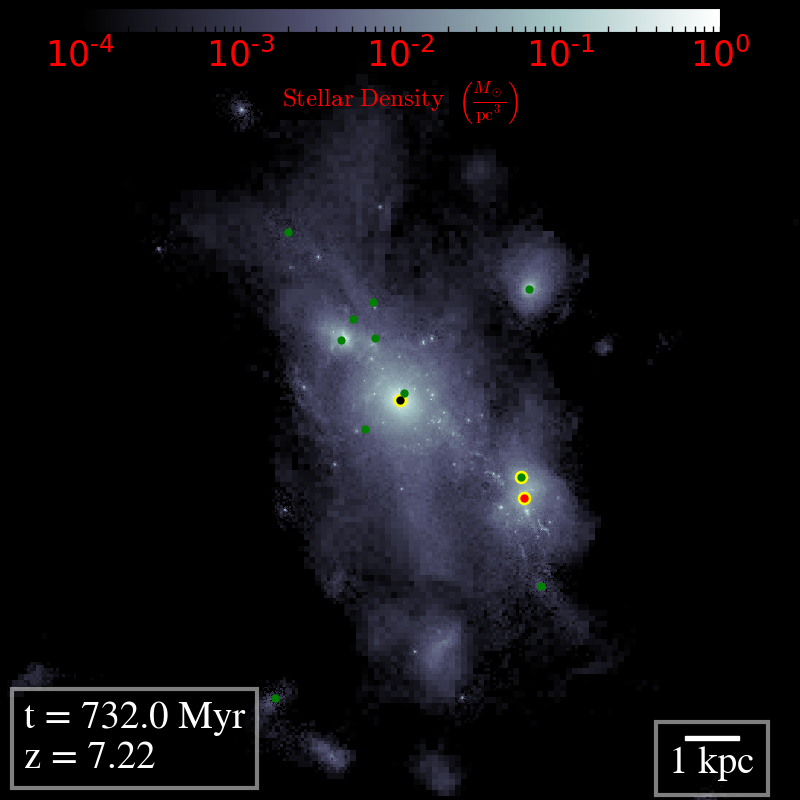}
\includegraphics[width=0.49\textwidth]{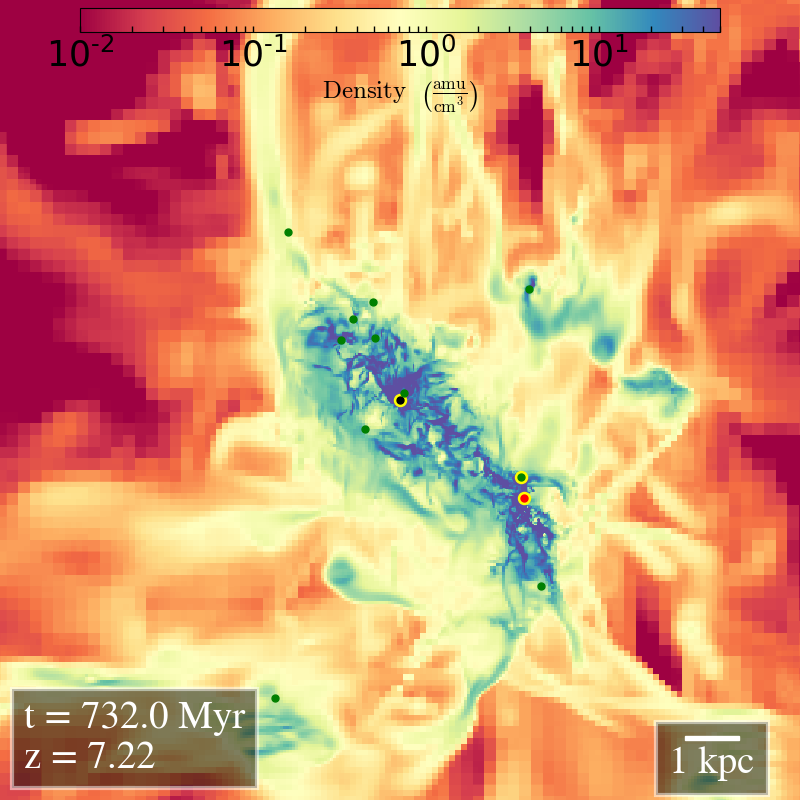} \\
\includegraphics[width=0.49\textwidth]{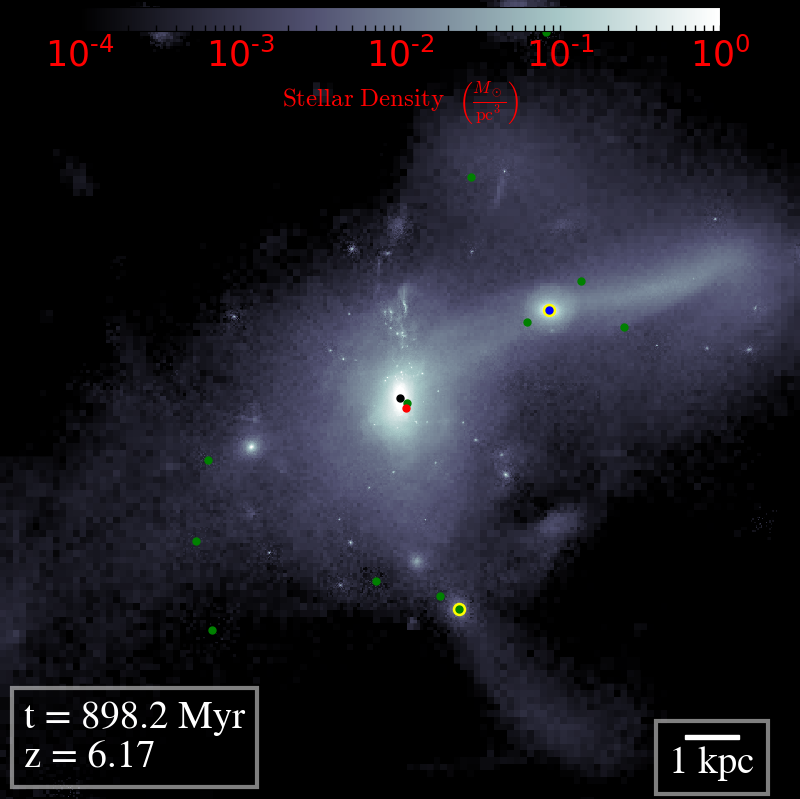}
\includegraphics[width=0.49\textwidth]{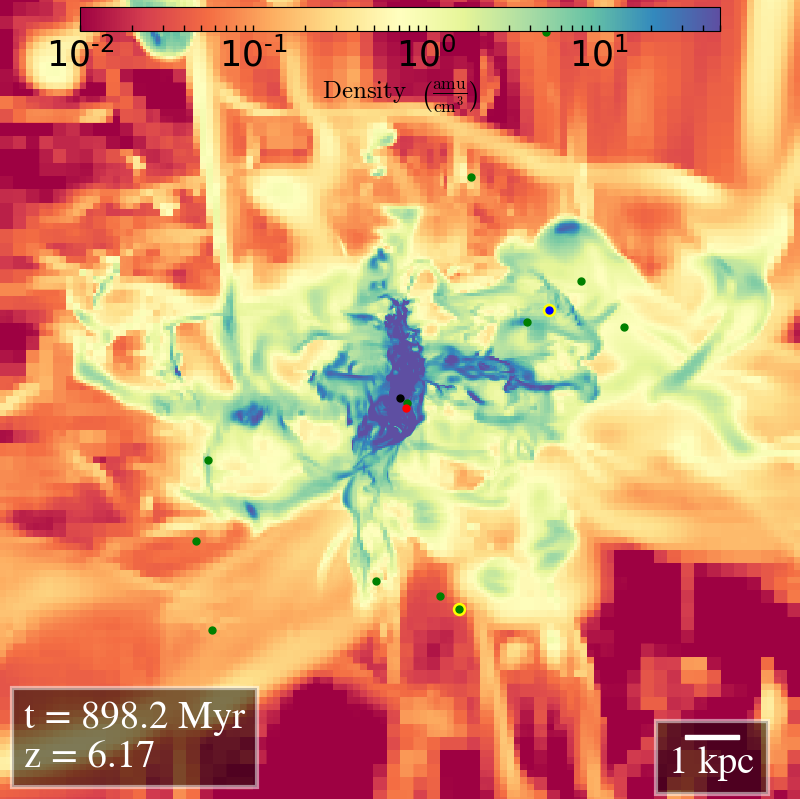}
\caption{\textbf{Left:} Stellar density projection. \textbf{Right:} gas--density--weighted gas density projection. In all cases, the images are centered on the main galaxy. \textbf{Top:} Moment at which the main galaxy undergoes a minor 1:10 merger, the satellite galaxy is in the bottom right of the main galaxy and contains the ``minor'' MBH (red dot). \textbf{Bottom:} Moment at which the main galaxy undergoes a major 1:4 merger, the satellite galaxy is on the top right of the main galaxy and contains the ``major'' MBH (blue dot). In all cases, we show all the MBHs (dots): the central MBH of the main galaxy (black), the minor MBH (red), the major MBH (blue) as well as all the other MBHs in the field of view (green). Finally, we indicate MBHs which have, at the time of the snapshot, a TDE rate larger than $10^{-5}\yr \mo$ with a yellow ring. The colors used for the 3 ``special'' MBHs are the same as in Fig.~\ref{fig:Gamma}.}
\label{fig:snapshots}
\end{figure*}

\subsection{Physics of black holes}
\label{sec:BHImplementation}

Our model for MBHs follows closely from \cite{Dubois_12}.

\subsubsection{Seeding}

MBHs are represented with sink particles, with an initial mass $M_{\bullet,\mathrm{seed}}=10^5\Msun$. They are formed in Jeans unstable cells containing enough gas to form the MBH, and with  ${\min(\rho_\star, \rho_\mathrm{gas})>100\amu\cm^{-3}}$, where $\rho_\star$ ($\rho_\mathrm{gas}$) corresponds to the stellar (gas) density in the cell. As this criterion formation is local, \ie we do use any halo finder to enforce MBH seeding in the exact center of halos/galaxies \cite[\eg][]{Vogelsberger_13}, this could result in multiple MBHs per galaxy. In order to avoid this, an exclusion radius of 50 kpc is used.

\subsubsection{Accretion}
\label{sec:BHAccretion}

Each MBHs are surrounded by massless cloud particles equally spaced by $\Delta x/2$ on a regular grid lattices within a sphere of radius $4\Delta x$ around the MBH. These cloud particles are used to measure the averaged gas quantities around the MBH. For instance, the mean gas density is obtained as:
\begin{eqnarray}
\tilde{\rho}_\mathrm{g} = \sum_{i \in \mathrm{cloud\, particles}} \rho_\mathrm{g, i} \exp\left( - \frac{r^2_i}{r^2_\bullet} \right) \, ,
\end{eqnarray}
where $\rho_\mathrm{g, i}$ is the gas density of the cell the cloud particle lies in and $r_i$ is the distance of the cloud particle to the MBH. $r_{\bullet }$ is defined as: 
\begin{eqnarray}
r_{\bullet } = 
\begin{cases}
\frac{\Delta x }{4} &\textrm{if } r_{B} <  \frac{\Delta x }{4}\\
 r_{B} & \textrm{if }   \frac{\Delta x }{4}< r_{B} < 2 \Delta x \\
2 \Delta x &\textrm{if }   2 \Delta x <  r_{B}
\end{cases} \, ,
\end{eqnarray}
where $r_{B}=\G M_\bullet / (c^2_s + v^2_{\bullet,\mathrm{g}})$ is the Bondi radius \citep{Bondi_52}; $c_s$  and $v_{\bullet,\mathrm{g}}$ are respectively the sound speed and relative velocity of the MBH with respect to the gas, in the cell the MBH lies in.

From these averaged quantities we can estimate the gas accretion rate $\dot{M}_{\bullet, \textrm{gas}}$, using the minimum between the Bondi and the Eddington \citep{Eddington_16} accretion rate:
\begin{eqnarray}
\dot{M}_B &=& \frac{4\pi \G^2 M^2_\bullet \tilde{\rho}_\mathrm{g}}{ (\tilde{c}^2_s + \tilde{v}^2_{\bullet,\mathrm{g}})^{3/2}} \label{eq:BondiAcc}\\
\dot{M}_\mathrm{Edd}  &=& \frac{4\pi \G \Mbh m_p}{\epsilon_r \sigma_T c} \, ,
\end{eqnarray}
where $m_p$ is the proton mass; $c$ is the speed of light; $\sigma_T$ is the Thompson cross-section and $\epsilon_r$ is classically fixed to 10\% as the spin is not followed in the simulation.

In addition, stellar accretion onto MBHs through TDEs, as described in \S\ref{sec:Implementation}, is used. In what follows we refer to the total MBH accretion rate as $\dot{M}_\bullet$.

\subsubsection{AGN feedback}
\label{sec:AGNfeedback}

Following accretion, between  $t$ and $t+\Delta t$, the energy released in the medium is 
\begin{eqnarray}
E_\mathrm{AGN} = \epsilon_r \epsilon_f \dot{M}_\bullet c^2 \Delta t \, ,
\end{eqnarray} 
where $\epsilon_f$ is the coupling efficiency, indicating how does the energy released couples with the gas and depends on the mode the MBH is in.

At high accretion rate (Eddington ratio, $\chi = \dot{M}_\bullet/\dot{M}_\mathrm{Edd} > 1\%$), the (thermal) energy is uniformly distributed in all cells within $r_\mathrm{AGN} = 4\Delta x$ from the MBH: this is the thermal mode. In this situation we set $\epsilon_f = 1.5\%$, lower than the value from \cite{Dubois_12} or \cite{Trebitsch_19}, but larger than \cite{Lupi_19} and similar to \cite{Capelo_15}.

At low accretion rate ($\chi < 1\%$) the (kinetic) energy is released through a cylindrical bipolar jet centered on the MBH, with radius/height $r_\mathrm{AGN}$ and direction parallel to the angular momentum of surrounding gas:
\begin{eqnarray}
\vv{L}_\mathrm{g}=\sum_{i\in \mathrm{cloud\, particles}} \rho_{\mathrm{g},i} \vv{r_i} \times \vv{u_i} \, ,
\end{eqnarray}
where $\vv{r_i}$ and $\vv{u_i}$ are respectively the distance and velocity relative to the MBH of the gas cell hosting the cloud particle $i$. The rate at which momentum is deposited depends on the radial distance $r$ to the axis of the cylinder:
\begin{eqnarray}
\dot{p}_\mathrm{Jet}(r)  = \psi(r) \eta \dot{M}_\bullet   \times \sqrt{\frac{2 \epsilon_r \epsilon_f}{ \eta}} c \, ,
\end{eqnarray}
where $\epsilon_f=100\%$ as in \cite{Dubois_12}; $\eta=100$ is the mass loading factor, corresponding the the enhancement of the mass due to swept up gas\footnote{Note that the speed of the jet is $10^4\kms$ with the parameter chosen, whereas in reality jets are relativistic. The difference is due to our lack of resolution (7 pc) and the jet should instead be considered as a wind.}, and:
\begin{eqnarray}
\psi(r) \propto \exp\left(- \frac{r^2}{r^2_\mathrm{AGN}} \right)
\end{eqnarray}
sums up to 1 over the whole cylinder.

\subsubsection{Dynamics}
\label{sec:Dynamics}

Contrary to many simulations where MBHs are anchored to the center of galaxies \citep[\eg][]{Vogelsberger_13}, we allow MBHs to freely move in the potential. Being massive, they suffer dynamical friction \citep{Chandrasekhar_43, BT_87, Tremmel_15}, some of which is unresolved due to lack of resolution \citep{Pfister_17}. For this reason, additional forces, in the opposite direction of the velocity of the MBH, are added to correct the dynamics.

Dynamical friction from stars/dark matter is detailed in \cite{Pfister_19a} \citep[using analytical work from][]{Chandrasekhar_43}, and dynamical friction from gas is detailed in \cite{Dubois_14a}  \citep[using analytical work from][]{Ostriker_99}. To our knowledge, \Ramses\, is currently the only code which physically treats both collisional and collisionless unresolved dynamical friction.

Finally, we stress that we have chosen a relatively massive MBH seed ($M_{\bullet, \mathrm{seed}} = 10^5\Msun > 10\, m^\textrm{part}_\star$), as such, these MBHs are not subject to spurious 2-body interactions and no additional correction is needed for the dynamics \citep{Pfister_19a}.
\subsubsection{Mergers}
\label{sec:Mergers}
When two MBHs get closer than $4\Delta x$, and if the gravitational energy of the binary is larger than the kinetic energy, \ie the binary would be bound in vacuum, MBHs are numerically merged. Note that this could lead to spurious mergers  \citep{Volonteri_20}, which we do not explore in this paper.

\subsection{Halos, galaxies, their history and some matching}
\label{sec:HaloFinderMatching}

We use \textsc{AdaptaHOP} \citep{Aubert_04} on dark matter (stellar) particles to detect gravitationally bound structures, \ie halos (galaxies), containing at least 50 particles. We then construct the history of halos (galaxies) using \textsc{TreeMaker} \citep{Tweed_09}, which match halos (galaxies) from one output to the other using the IDs of particles forming the structures.

We then match galaxies to halos, selecting the closest galaxy in position. As the zoom has been made on a particular halo, which is the most massive one unpolluted, \ie containing only high resolution dark matter particles, the galaxy of this halo is the ``main'' galaxy. Galaxies which are identified and are matched to other unpolluted halos are called ``satellite'' galaxies.

Finally, we match MBHs to galaxies. A MBH is assumed to belong to a galaxy if it is within the effective radius of the galaxy (see definition in Appendix~\ref{sec:EffectiveRadiusOfGalaxies}), and the closest to the center is the central MBH of this galaxy. If a MBH can be associated to many galaxies, we assign the MBH the most massive galaxy. In what follows, we refer to the ``central'' MBH as the central MBH of the main galaxy at the end of our simulation (at $z\sim6$).

\begin{figure}
\includegraphics[width=\columnwidth]{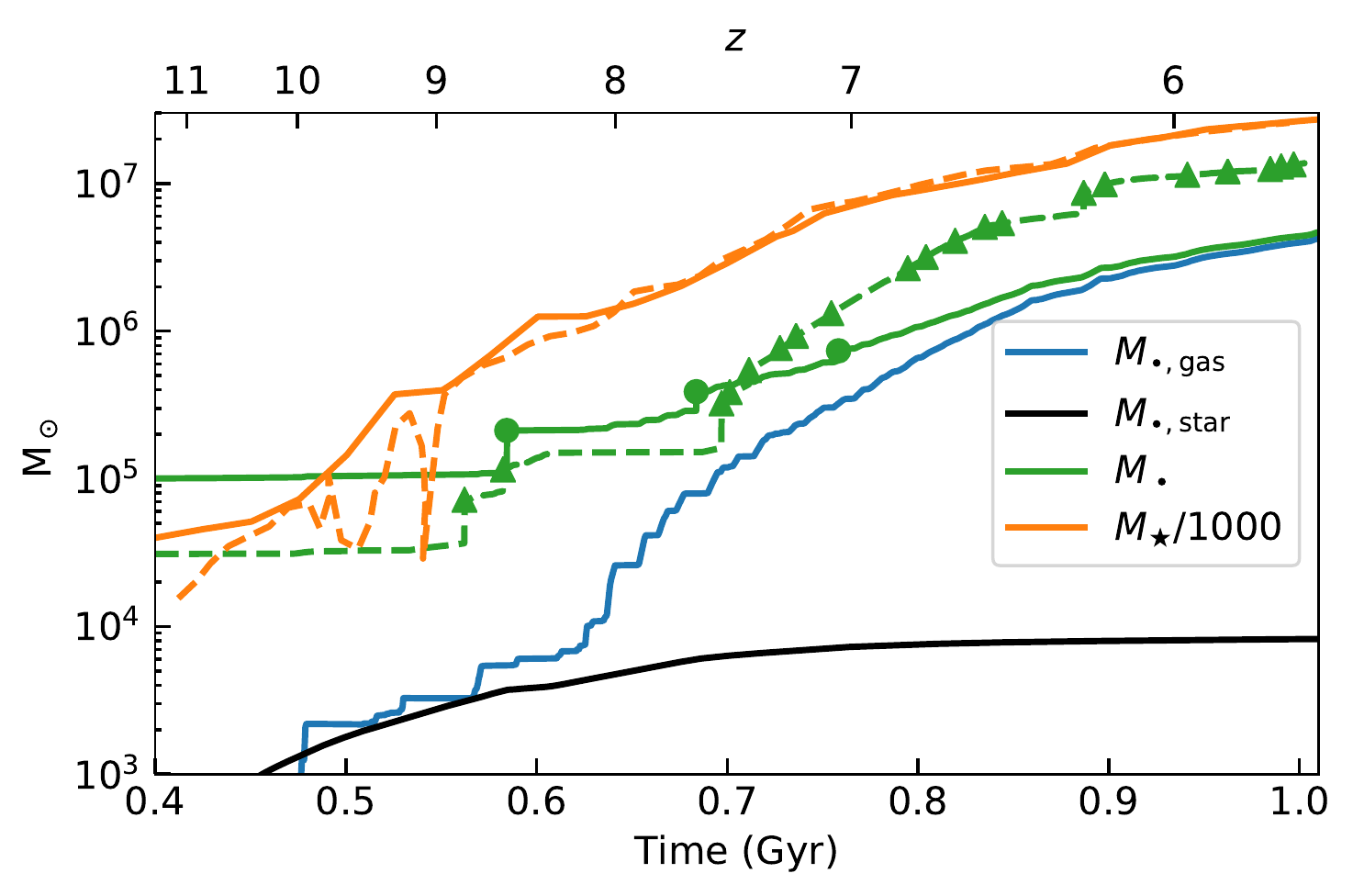}
\caption{Masses of the central MBH (green) and the main galaxy (orange) as a function of time in our simulation (solid lines) and in the simulation of {\protect\cite{Trebitsch_19}} (dashed lines). We also show the accreted mass of gas (blue) and of stars following TDEs (black). MBH mergers are indicated with dots (this work) or triangles {\protect\citep{Trebitsch_19}}. In the end, the total contribution of accretion following TDEs is negligible, except at early time, where accretion from TDEs and gas is similar.}
\label{fig:masses}
\end{figure}

In Fig.~\ref{fig:snapshots} we show the stellar (gas) density projection of the main galaxy during a minor 1:10 and a major 1:4 merger. We indicate MBHs with dots: the central MBH (black), the central MBH of the satellite galaxy of the minor merger (the ``minor'' MBH in red), the central MBH of the satellite galaxy of the major merger (the ``major'' MBH in blue) as well as all the other MBHs in the field of view (green). Finally, we indicate MBHs which have, at the time of the snapshot, a TDE rate larger than $10^{-5}\yr \mo$ with a yellow ring.

\section{Results}
\label{sec:Results}

\subsection{Global properties}

As we have used the exact same initial conditions as \cite{Trebitsch_19}, we can make a fair comparison between the global properties of the two simulations, keeping in mind that details may vary, as some parameters are not exactly the same \citep[particle stellar mass, seed mass of MBHs, use of boost for gas dynamical friction, absence of TDEs etc... see \S\ref{sec:NumericalSetUp} and][]{Trebitsch_19}.

We show in Fig.~\ref{fig:masses} the mass of the main galaxy (orange) as a function of time in our simulation (solid line) and in the simulation of  \cite{Trebitsch_19} (dashed line). Apart from minor differences at early time, as soon as the galaxy is well settled with a mass larger than $10^9\Msun$, its mass is independent of the detailed parameters of the simulation. 

On the same Figure, we show the mass of the central MBH (green) in the two simulations, as well as the moments at which the central MBH undergoes a MBH merger (markers). The final masses, which differ by a factor of 3, match remarkably well considering that (in unranked order) \textit{(i)} the initial MBH seed masses are different; \textit{(ii)} \cite{Trebitsch_19} uses a boost for gas dynamical friction, ``encouraging'' the MBH to remain in gas dense regions and reducing its relative velocity to surrounding gas, enhancing the accretion rate (which scales as the density and the inverse cubic of the relative velocity,  see Eq.\eqref{eq:BondiAcc}), sometime by orders of magnitude; \textit{(iii)} the number of mergers, and the total ``accreted'' mass through mergers greatly differ: 3 mergers in our simulation corresponding to 6\% of the final mass, and 20 mergers in \cite{Trebitsch_19} corresponding to 24\% of the final mass (this is likely to be related to \textit{(i)} and \textit{(ii)}, but we leave this for future investigations, as we are interested in the TDE rate in this paper); \textit{(iv)} \cite{Trebitsch_19} do not include MBH growth through TDEs; and \textit{(v)} the AGN feedback coupling efficiency in the thermal mode differs by a factor of 10 in the two simulations.

On the same Figure, we show the mass accreted through gas (blue), and through stars following TDEs (black). As the total contribution of TDEs is only $10^4\Msun$ out of the $5\times10^6\Msun$ of the MBH final mass, this suggests that the difference between \cite{Trebitsch_19} and our simulation is not due to \textit{(iv)}, and including TDEs is not mandatory to properly estimate the final mass of the MBH. However, at early time, when the MBH is lighter than $\sim 5\times10^5\Msun$, the contribution of stars appear to be similar to that of gas. 

\begin{figure}
\includegraphics[width=\columnwidth]{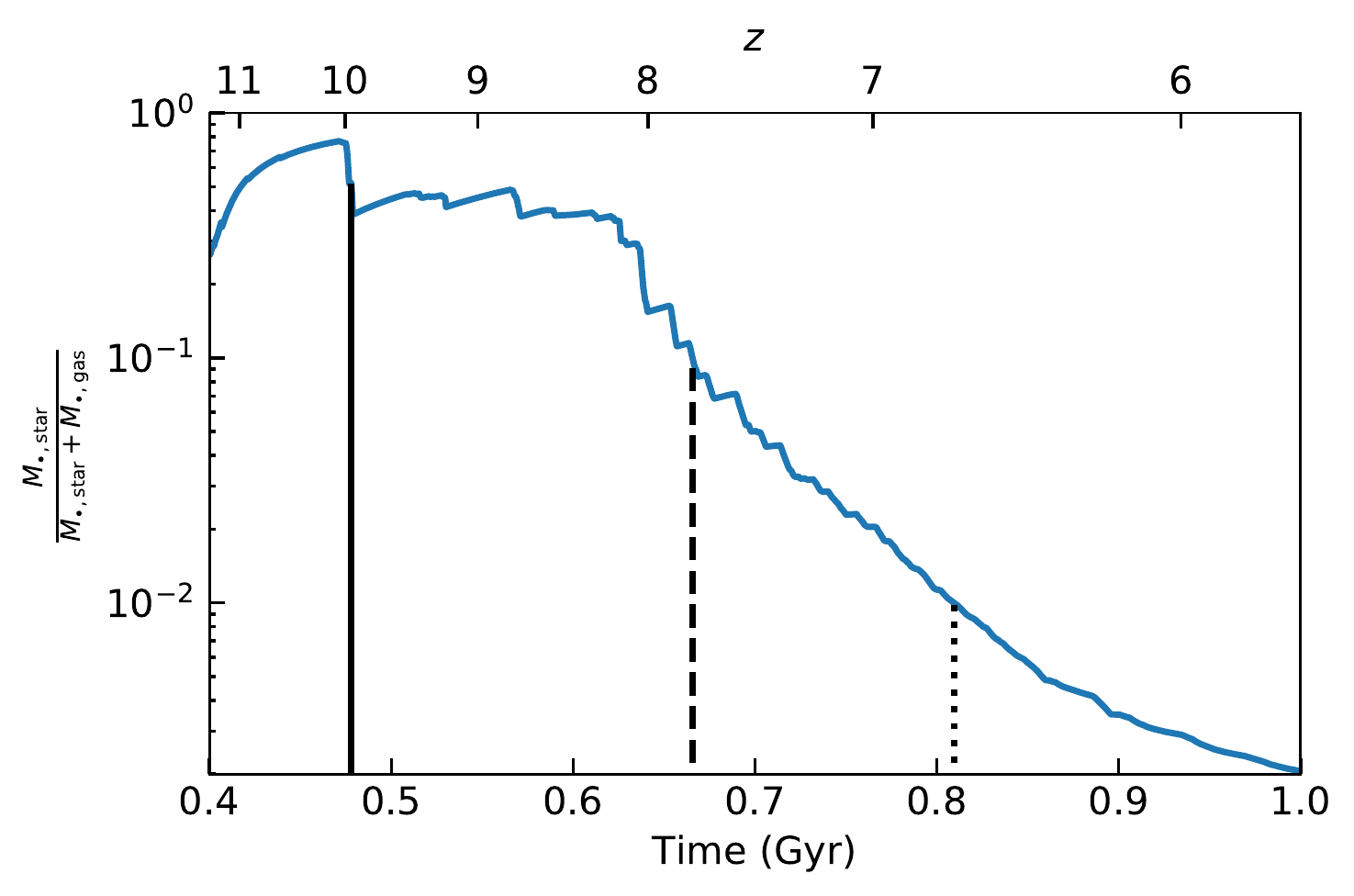}
\caption{Ratio of the mass accreted through TDEs ($M_\mathrm{\bullet,\, star}$) with the total mass accreted from gas and stars ($M_\mathrm{\bullet,\, star}+M_\mathrm{\bullet,\, gas}$), as a function of time, for the central MBH. We also indicate the last time at which fraction is larger than 50\% (solid black line), 10\% (dashed black line) and 1\% (dotted black line). In the end, about $0.1\%$ of the mass is gained from TDEs, and their contribution is negligible for massive MBHs. However, during the first 300~Myr, their contribution is larger than 10\%.}
\label{fig:accfraction}
\end{figure}

In Fig.~\ref{fig:accfraction} we show the fraction of mass accreted through star as a function of time. At early time, TDEs and their subsequent stellar accretion have a significant contribution to the growth of the MBH. Indeed, more than 10\% of the accreted mass of the central MBH is coming from stars during the first 300~Myr of its life, until its mass is larger than $\sim 5\times10^5\Msun$ and the MBH is massive enough to accrete at about the Eddington rate and mostly grow through gas accretion. Unfortunately, for numerical reasons (see \S\ref{sec:Dynamics}), we could not decrease the seed mass of the MBH and study the earlier growth of intermediate mass MBHs through TDEs.  We note however that this is in principle doable with the models described in \S\ref{sec:Implementation}, at the cost of globally increasing the resolution of the simulation.

\begin{figure}
\includegraphics[width=\columnwidth]{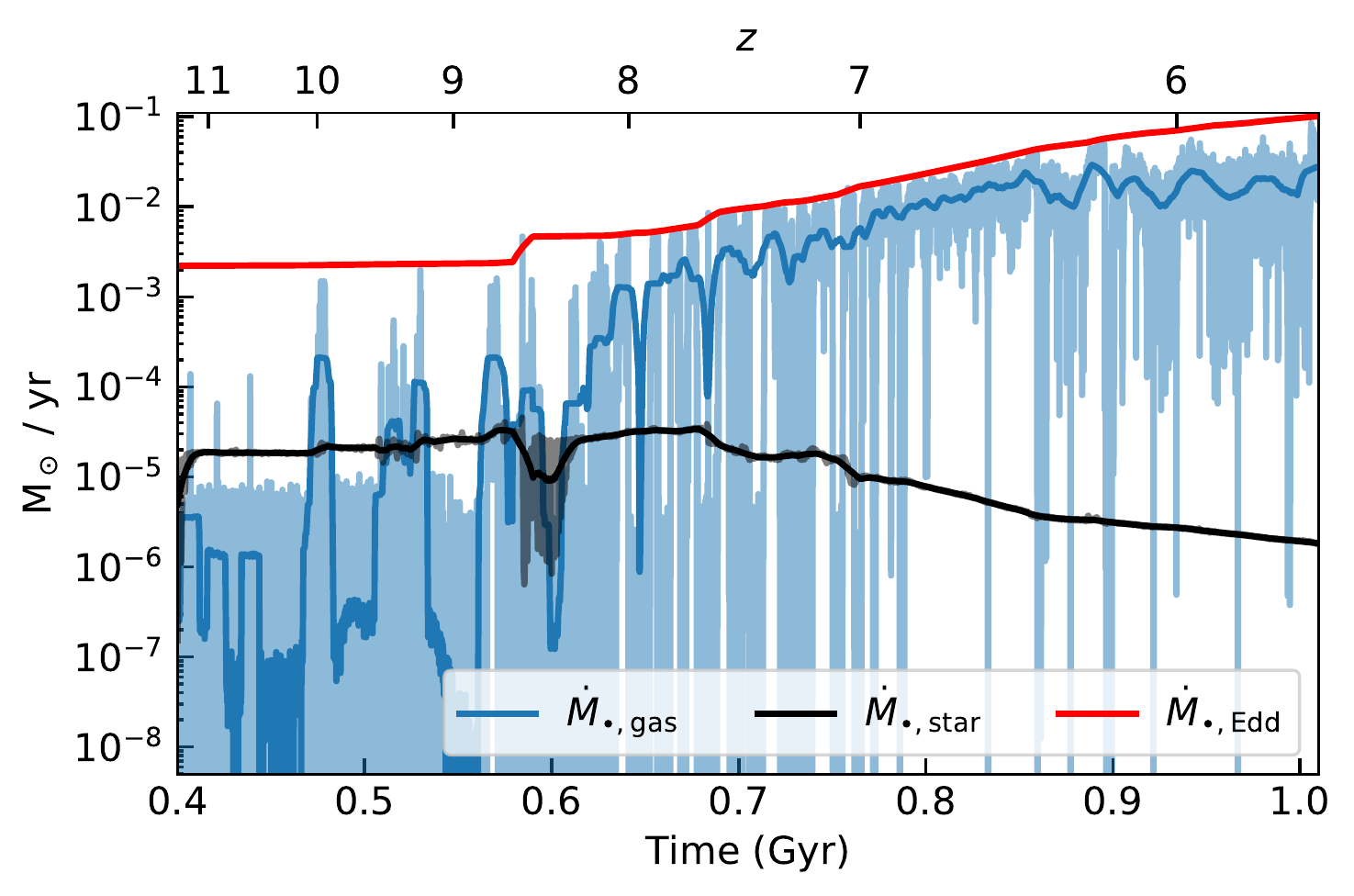}
\caption{Gas (blue), stellar (black) and Eddington (red) accretion rate of the central MBH. Light colors are direct outputs of the simulation (every $50 \kyr$) and dark colors are averaged with a 10 Myr window. The stellar accretion rate is strikingly smoother than the gas accretion rate, although the final contribution of the latter is larger (see Fig.~\ref{fig:masses} and Fig.~\ref{fig:accfraction}).}
\label{fig:accretionrate}
\end{figure}

In Fig.~\ref{fig:accretionrate}, we show the gas (blue), stellar (black) and Eddington (red) accretion rate of the central MBH averaged on different timescales (light color, $50 \kyr$; dark color, 10 Myr). Although they are shown with the same frequency, the stellar accretion rate is smoother than gas accretion rate. The reason is twofold: \textit{(i)} the stellar density is \textit{spatially} smoother than the gas density (see Fig.~\ref{fig:snapshots} for projections maps), therefore changes in the MBH position will change the gas density (and MBH gas accretion), leaving the stellar density (and the MBH stellar accretion) unchanged; and \textit{(ii)} stars are not subject to feedback while gas is, so at a given spatial position, the stellar density is \textit{temporally} smoother than the gas density \citep{Prieto_17}. More quantitatively, we simply estimate smoothness of a quantity $\overline{u}$ as the time average of the relative variation throughout the simulation:
\begin{eqnarray}
\overline{u} = \left\langle \left| \frac{\Delta u}{u} \right| \right\rangle \, ,
\end{eqnarray}
where $\Delta u$ is the variation of the quantity $u$ between two consecutive timesteps (about 50 kyr); $u$ is the mean value of the quantity $u$ on two consecutive timesteps and  $\left\langle . \right\rangle$ indicates an average over the duration of the simulation. We find that ${( \overline{\dot{M}}_{\bullet,\mathrm{star}}, \overline{\dot{M}}_{\bullet,\mathrm{gas}}, \overline{\rho}_{0,\mathcal{S}}, \overline{\tilde{\rho}}_\mathrm{g}) = (13\%, 66\%, 1\%, 60\%)}$. The relative variation in the gas accretion ($\overline{\dot{M}}_{\bullet,\mathrm{gas}}$) reproduces well the relative variations of the the gas density in the vicinity of the MBH ($\overline{\tilde{\rho}}_\mathrm{g}$). While the relative variations of the stellar accretion ($\overline{\dot{M}}_{\bullet,\mathrm{star}}$) does not reproduce as well the relative variations of the stellar density around the MBH ($\overline{\rho}_{0,\mathcal{S}}$), we recall that, contrary to gas accretion, stellar accretion does not scales directly linearly with the stellar density around the MBH.

This confirms however that the rapidly (slowly) varying gas (stellar) density around the MBH results in a rapidly (slowly) varying gas (stellar) accretion. We note that, because it is much smoother, at any times (hence MBH masses), the accretion rate following TDEs can be orders of magnitude larger than the gas accretion rate. This suggests that, at any time, it is possible that the properties of the emitting MBH are those of a MBH accreting stars only. If the composition of stars differ from the composition of surrounding gas (\eg stars have a higher nitrogen to carbon abundance), this confirms that, at any time, nitrogen rich quasar could be due to TDEs \citep{Kochanek_16, Liu_18}.

To summarize, stellar accretion due to TDE is smoother than gas accretion, simply due to that the stellar density in the vicinity of the MBH is smoother than the gas density, and stellar accretion can be much larger than gas accretion at all time. However, in the end, growth through TDEs is efficient only for MBHs with a mass lower than $5\times10^5\Msun$, more massive MBHs mostly grow through gas accretion and the final TDEs contribution is negligible.

\subsection{TDE rate}
\label{sec:TDErate}

\begin{figure}
\includegraphics[width=\columnwidth]{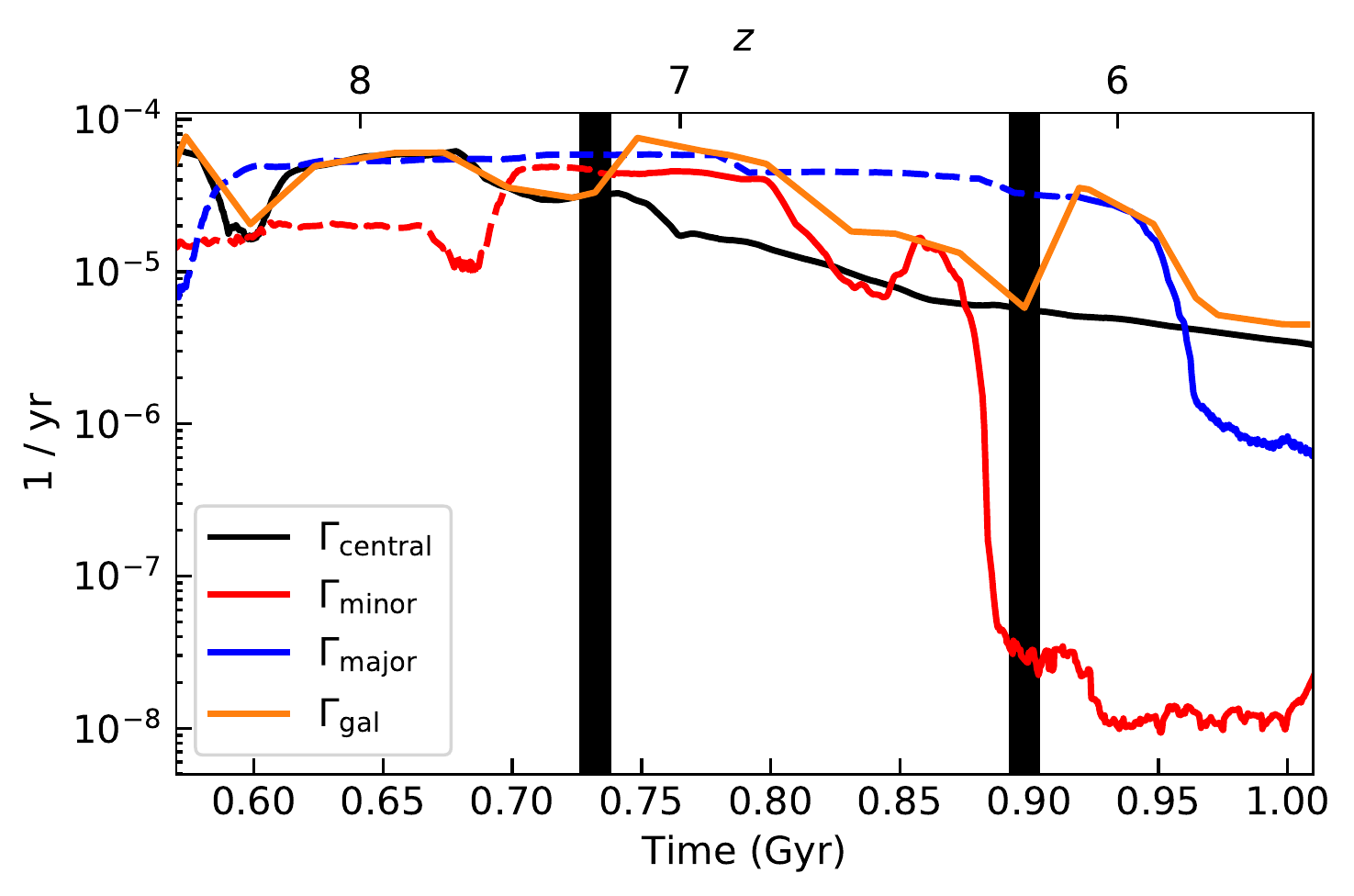}
\caption{TDE rate as a function of time of the central MBH of the main galaxy (black), of the central MBH of the secondary galaxy during the 1:10 minor merger (red), and of the central MBH of the secondary galaxy during the 1:4 major merger (blue). These MBHs are shown with the same colors as in Fig.~\ref{fig:snapshots}. We also show the total TDE rate of the main galaxy (orange). All TDE rates are averaged with a 10~Myr window. The two thick black vertical areas indicate two moments at which the main galaxy undergoes a merger, and for which we show stellar/gas density projection maps in Fig.~{\protect\ref{fig:snapshots}}. When MBHs are not in the main galaxy, we indicate their evolution with dashed lines, their subsequent evolution following the galaxy merger is marked with solid lines. We find a clear enhancement of $\sim 1$ order of magnitude of the total TDE rate of the main galaxy during mergers, however, the enhancement does not occur on the central MBH.}
\label{fig:TDErate}
\label{fig:Gamma}
\end{figure}

Our simulation allows us to estimate the TDE rate of every MBHs as a function of time. Since we also know which MBHs belong to the main galaxy, we can estimate the total TDE rate of the galaxy as:
\begin{eqnarray}
\Gamma_\mathrm{gal} = \sum_{i\in\mathrm{BHs\, in\, the\, main\, galaxy}} \left\langle\Gamma_i\right\rangle_{10\Myr}\, ,
\end{eqnarray}
where $\Gamma_i$ is the TDE rate of MBH $i$ and $\left\langle . \right\rangle_{10\Myr}$ indicates an average over a 10~Myr window (our results are unchanged with a 5 or 50~Myr window).

In what follows we will focus on the three ``special'' MBHs presented in \S\ref{sec:HaloFinderMatching} and Fig.~\ref{fig:snapshots}: the central MBH of the main galaxy (the ``central'' MBH in black), the central MBH of the satellite galaxy of the minor merger (the ``minor'' MBH in red) and the central MBH of the satellite galaxy of the major merger (the ``major'' MBH in blue).

\subsubsection{TDE rate during mergers}
\label{sec:TDErateDuringMergers}

We show in Fig.~\ref{fig:Gamma} the TDE rates of the 3 MBHs $\Gamma_\mathrm{central}$/$\Gamma_\mathrm{minor}$/$\Gamma_\mathrm{major}$ in black/red/blue (colors are the same as the dots representing these MBHs in Fig.~\ref{fig:snapshots}) as well as the total TDE rate of the galaxy $\Gamma_\mathrm{gal}$ (orange) as a function of time. The minor and major galaxy mergers shown in Fig.~\ref{fig:snapshots} are indicated with thick vertical black areas. When the MBHs of the satellite galaxies are not in the main galaxy (they are brought by the galaxy merger), we indicate their evolution with a dashed line.

The total TDE rate of the galaxy (orange) is few $10^{-5}\yr\mo$. This value is in good agreement with local estimates \citep{Donley_02, Gezari_08, VanVelzen_14, Holoien_16, Blagorodnova_17, Auchettl_18, VanVelzen_18} but already in place at $z\gtrsim 6$. We recall that, by construction only one fairly massive galaxy is studied here (this is a zoom-in simulation), and a more statistical analysis should be performed, but this suggests that some galaxies could already have a well established TDE rate of few $10^{-5}\yr\mo$ at $z\sim6$ when the universe is 1~Gyr.

Initially, the total TDE rate of the galaxy (orange) is similar to that of the central MBH (black), \ie the TDE rate of the galaxy is dominated by TDEs occuring on the central MBH. However, MBHs brought by successive mergers (all the dots but the black one in Fig.~\ref{fig:snapshots}), which can take very long time to sink toward the center of the galaxy through dynamical friction \citep{Pfister_19a}, also contribute to the total TDE rate of the galaxy, sometime dominating it.

For instance, during the first minor merger we consider (at $t=0.73\Gyr$), the MBH of the satellite galaxy (the minor MBH in red), which has a high TDE rate ($4\times10^{-5}\yr\mo$) penetrates the main galaxy, resulting in an enhancement the total TDE rate. This high TDE rate around the minor MBH is due to a merger induced nuclear starburst at $t=0.70\Gyr$, time at which the star formation rate within $4\Delta x = 28\pc$ from the minor MBH is enhanced by 30. This picture is in agreement with previous theoretical results who find that mergers trigger nuclear starbursts, enhancing the TDE rate \citep{Pfister_19b}.

During the second major merger we study (at $t=0.90\Gyr$) the major MBH (blue) penetrates the main galaxy and completely dominates the rate. The picture here is however different than that of the first merger, as the TDE rate around this major MBH is not enhanced \textit{per se}: it was of $5\times10^{-5}\yr\mo$ since $t=0.60\Gyr$. Instead, the major MBH penetrates the main galaxy while being surrounded by an already dense stellar cusp (see bottom left panel of Fig.~\ref{fig:snapshots}), therefore its already high TDE rate is not affected.

Overall, we find that during the two mergers we discussed, the TDE rate is enhanced by 1 order of magnitude during about 100~Myr. This enhancement is due to a nuclear starburst for the first minor merger, and to that a MBH with a well established stellar cusp enters the main galaxy for the second major merger. Other processes resulting in an enhancement of the TDE rate could happen during mergers: dynamical effects in dry mergers \citep{LiShuo_17}; or simply a MBH on an eccentric orbit periodically crossing the dense center of the main galaxy. We did not find such examples in our simulation.

Finally, we note that the TDE rate of the central galaxy is dominated by off-centered MBHs during about 200~Myr out of the 1~Gyr our simulation lasts, suggesting that during up to 20\% of the time, the TDE rate could be dominated by off-centered TDEs. While surveys designed to find TDEs \citep[\eg][]{VanVelzen_20} usually look for central TDEs to exclude most supernovae, blind surveys may already have observed off-centered TDEs \citep{Lin_18, Margutti_19}.

To summarize, we find that, for some galaxies at least, the TDE rate at $z>6$ could already be similar to the one at  $z=0$. We also confirm that the TDE rate is globally enhanced by about 1 order of magnitude during 100~Myr around mergers, but not necessarily for the central MBH of the main galaxy. MBHs brought by successive mergers could see their TDE rate larger than the one of the central MBH, and actually dominate the total TDE rate of the galaxy, resulting in fairly frequent ($\sim 20\%$ of the time in our simulation) off-centered TDEs.

\subsubsection{TDE rate in AGNs}
\label{sec:TDErateInAGNs}

\begin{figure}
\includegraphics[width=\columnwidth]{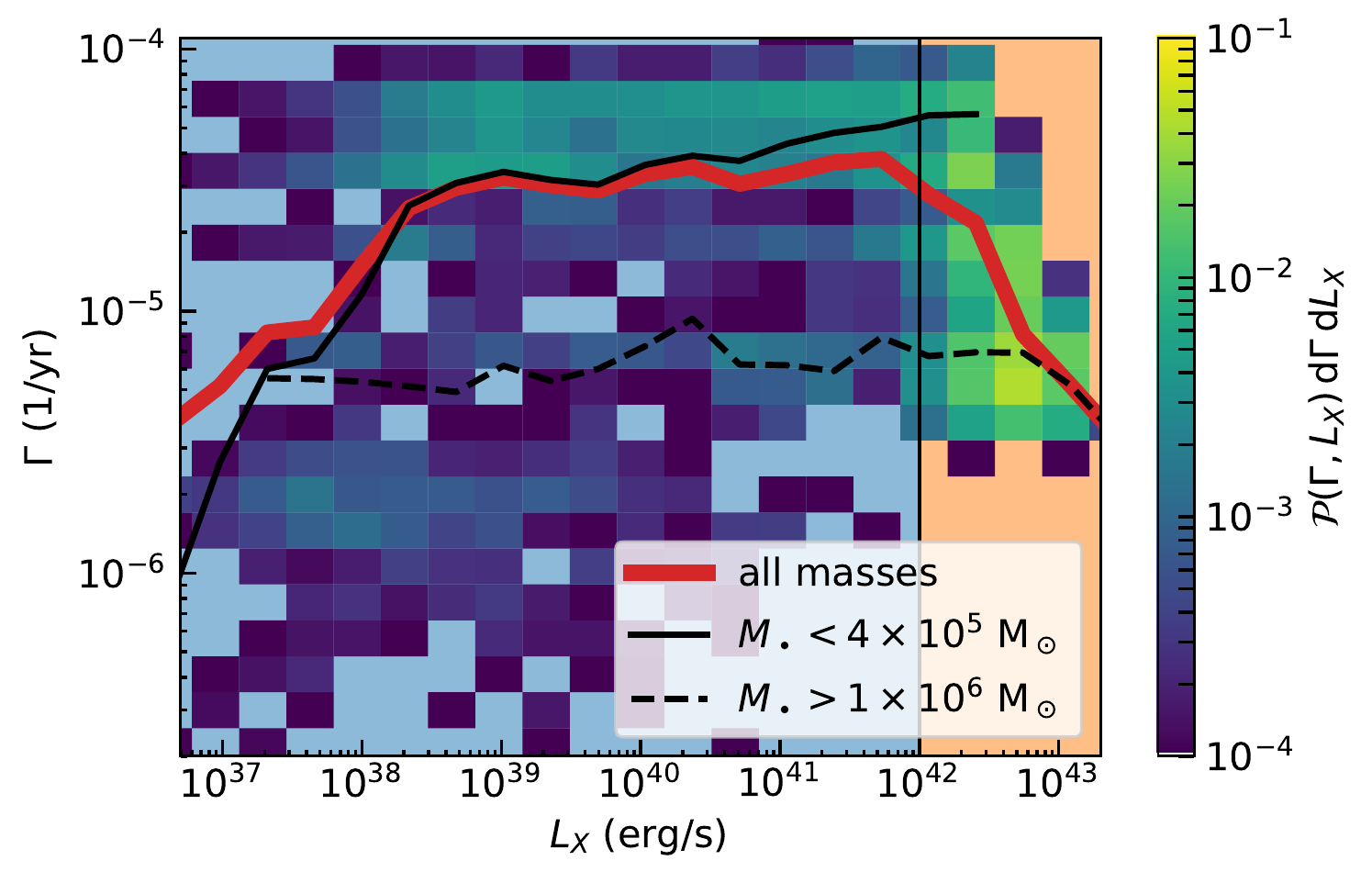}
\caption{Fraction of time spent at a a given TDE rate ($\Gamma$) and X-ray luminosity ($L_X$). We show the mean TDE rate at fixed $L_X$ for all MBH masses (red), for light MBHs (solid black line) and for more massive MBHs (dashed black line). For all MBH masses, there is a decrease in the TDE rate in AGNs. This is due to that AGNs are usually powered by more massive MBHs, with a lower TDE rate. At fixed mass, the TDE rate is independent of the X-ray luminosity.}
\label{fig:FractionTimeVSGammaLx}
\end{figure}

As AGNs and TDEs share the properties of having strong variability and being quite luminous, it is challenging to detect TDEs in AGNs using standard methods and, in general, AGNs are excluded from searches of TDEs \citep[\eg][]{VanVelzen_20}. For these reasons, few candidates of TDEs in AGNs have been suggested \citep[\eg][]{Blanchard_17}, and it is currently difficult to constrain the TDE rate in AGNs from observations. Nonetheless, several groups suggest that up to 10\% of AGNs are powered by TDEs \citep{Milosavljevic_06, Merloni_12}. With our simulation, we can directly test what is the TDE rate when the galaxy has an AGN.

First, we have to define \textit{when} the main has an AGN. We follow \cite{Brightman_11} (see \S3.6 of their paper) and define the central MBH as an AGN if the X-ray luminosity in the $2-10 \keV$ band of the central MBH, $L_X$, is larger than $10^{42}\erg \s^{-1}$. To this purpose, we use the following bolometric correction \citep{Hopkins_07, ShenXuejian_20}:
\begin{eqnarray}
L_X & = & \frac{L_\mathrm{bol}}{k} \\
L_\mathrm{bol} &=& \frac{\epsilon_r}{1-\epsilon_r} \dot{M}_{\bullet,\mathrm{gas}} c^2 \\
k &=& 10.83  \left(\frac{L_\mathrm{bol}}{10^{10}L_\odot}\right)^{0.28} + 6.08\left(\frac{L_\mathrm{bol}}{10^{10}L_\odot}\right)^{-0.020} \, .
\end{eqnarray}
We \textit{exclude} here the stellar accretion when computing $L_\mathrm{bol}$. The reason is that \textit{including} stellar accretion would result in an X-ray background: the central MBH is constantly accreting stars at about $10^{-5}\Msun\yr\mo$, and taking into account stellar accretion would result in constant minimum X-ray luminosity of $\sim 10^{40}\erg\s\mo$. This artifact comes from our poor ($\sim 50\kyr$) temporal resolution: in reality, TDEs occurs on $\sim \yr$ timescale with much brighter luminosity \citep{Auchettl_17}, therefore do not produce this unphysical X-ray background. In other words, because the number of TDEs during one timestep is small ($\Gamma\Delta t \lesssim 1$), if we were to observe the galaxy during one timestep, the fraction of time during which the luminosity would be the one of a TDE would be very small ($\sim \Gamma \times 1\yr \sim 10^{-5}$ for a typical duration of 1 yr), and at much brighter $L_X$. We stress here that we do not pretend to capture the details of the luminosity curve to differentiate between AGNs and TDEs: both our spatial and temporal resolution are far too poor. Our goal here is to know what would be the typical TDE rate in an AGN. Note that we also exclude all the wandering MBHs of the main galaxy, which could also produce X-rays. We did so because none of them has an accretion rate similar to that of the central MBH: the second most massive MBH is only $5\times10^5\Msun$ (it is the major blue MBH from \S\ref{sec:TDErateDuringMergers}).

From $\Gamma$ and $L_X$ at all times, we can compute the joint distribution $\mathcal{P}(\Gamma,L_X)$, such that $\mathcal{P}\d \Gamma \d L_X$ corresponds to the fraction of time spent a $L_X$ and $\Gamma$, as:
\begin{eqnarray}
 \mathcal{P} \d \Gamma  \d L_X = \frac{ \sum \Delta t_i }{ \tau_\bullet} \, ,
\end{eqnarray}
where $i$ corresponds to timesteps during which the X-ray luminosity and TDE rate are respectively in $[L_X,L_X+\d L_X]$ and $[\Gamma,\Gamma+\d \Gamma]$; $\Delta t_i$ is the duration of these timestep and $\tau_\bullet\sim0.76\Gyr$ is the time during which the MBH is followed in the simulation.

We show $\mathcal{P}\d \Gamma \d L_X$ in Fig.~\ref{fig:FractionTimeVSGammaLx}. We find a large scatter, suggesting no clear relations between TDE rate and X-ray luminosity. We compute the mean TDE rate at fixed $L_X$ (red line):
\begin{eqnarray}
\tilde{\Gamma}(L_X) = \left( \int_\Gamma \mathcal{P} \Gamma \d \Gamma \right) \biggm/ \left( \int_{\Gamma} \mathcal{P} \d \Gamma \right) \, .
\end{eqnarray}
On average, the TDE rate increases with $L_X$ until $10^{39}\erg \s\mo$ where it plateaus at few $10^{-5}\yr\mo$. When $L_X$ reaches $10^{42}\erg \s\mo$ and the MBH is classified as an AGN \citep{Brightman_11}, the TDE rate starts decreasing, suggesting that the TDE rate is lower in AGNs. However, we recall that the TDE rate is lower for more massive MBHs \citep{Wang_04}, and that more massive MBHs can shine more (assuming their luminosity is a fraction of the Eddington luminosity). Therefore it could be that this lower TDE rate in AGNs is simply due to that MBHs in AGNs are usually more massive. 

To test this, we split the simulation in two sub-samples: when the MBH is less massive than $4\times 10^5\Msun$ (${t<0.69\Gyr}$; ${\tau_\bullet\sim0.44\Gyr}$), and when it is more massive than ${10^6\Msun}$ (${t>0.79\Gyr}$; ${\tau_\bullet\sim0.22\Gyr}$)\footnote{The third part, when the mass of the MBH is in between $4\times 10^5\Msun$ and $10^6\Msun$ is excluded to avoid spurious results due to arbitrary transition.}. We then recompute $\tilde{\Gamma}$ for these two sub-samples (black lines). Regardless of the X-ray luminosity, the TDE rate is larger for lighter MBHs, in agreement with \cite{Wang_04}. Regarding the enhancement, or not, of the rate in AGNs, we find that, as long as $L_X > 10^{38}\erg \s\mo$, the TDE rate is fairly constant at all $L_X$, confirming that the lower TDE rate in AGNs is due to more massive MBHs.

To summarize, our simulation suggests that, at fixed MBH mass, there is no enhancement of the TDE rate in AGNs. However, in general, the TDE rate should be lower in AGNs simply because AGNs are powered by massive MBHs, for which the TDE rate is lower.

\section{Conclusions}
\label{sec:Conclusions}

We have developed a physically motivated subgrid model to include stellar accretion on MBHs and TDEs in cosmological simulations, and we have performed a cosmological zoom simulation of a $3\times 10^{10}\Msun$ galaxy at $z \sim 6$. Our main findings are the following:
\begin{enumerate}
\item Overall, TDEs and stellar accretion do not contribute much to the growth of MBHs, in our particular case only 0.2\% of the final mass comes from stars. However, TDEs are particularly efficient in growing MBHs in their early life, when they are lighter than $\sim 5\times 10^5 \Msun$, with more than 10\% of the total accreted mass coming from stars during the first 300 Myr. We stress that this value could be underestimated as the minimum MBH mass allowed in our simulation is $10^{5}\Msun$, and that the TDE rate increases with decreasing mass. All this suggests that accretion following TDEs is a promissing channel to rapidly grow light MBHs.
\item Stellar accretion is much smoother than gas accretion, this results naturally from the stellar density being temporally and spatially smoother than the gas density. At any time, the gas accretion rate can be orders of magnitude lower or higher than the stellar accretion rate.
\item When a galaxy merger occurs, the global TDE rate in a galaxy can be enhanced by up to 1~order of magnitude during 100~Myr. This enhancement occurs on the central MBH of the satellite galaxy and it is caused by a nuclear starburst or a MBH entering the main galaxy with a dense stellar cusp (hence with a high TDE rate).
\item As galaxy mergers bring many MBHs which may take a long time to sink toward the center of the main galaxy, the amount of off-centre TDEs could be fairly high. In our simulation, the TDE rate of the main galaxy is dominated by off-centre TDEs during 20\% of the time.
\item Some galaxies with mass comparable to that of the Milky Way today could already have a well established TDE rate of ${10^{-5}-10^{-4}\yr\mo}$, comparable with local estimates, at $z>6$.
\item At fixed MBH mass, the TDE rate is independent of the X-ray luminosity of the central MBH, and no enhancement is expected in AGNs. However, since luminous AGN are powered by MBHs with mass $>10^6 \Msun$ and the TDE rate decreases as $M_\bullet$ increases, for a  population of AGNs the TDE rate is expected to be $<10^{-5}\yr\mo$.
\end{enumerate}

This is the first study of TDEs and their evolution over cosmic time using cosmological hydrodynamic simulations. While only one galaxy has been studied in this analysis, we are planning to run a cosmological volume in order to increase the statistical validity of our investigation and explore how stellar accretion and TDEs depend on the environment and properties of their galaxies.

\section*{Acknowledgments}
HP is indebted to the Danish National Research Foundation (DNRF132) and the Hong Kong government (GRF grant HKU27305119) for support. KAA and ERR are supported by the Danish National Research Foundation (DNRF132). Parts of this research were supported by the Australian Research Council Centre of Excellence for All Sky Astrophysics in 3 Dimensions (ASTRO 3D), through project number CE170100013. MT is supported by Deutsche Forschungsgemeinschaft (DFG, German Research Foundation) under Germany's Excellence Strategy EXC-2181/1 - 390900948 (the Heidelberg STRUCTURES Cluster of Excellence). The authors thank the Yukawa Institute for Theoretical Physics at Kyoto University. Discussions during the YITP workshop YITP-T-19-07 on International Molecule-type Workshop "Tidal Disruption Events: General Relativistic Transients" were useful to complete this work. This work was made possible with the access to the HPC resources of CINES under allocations DARK n\textdegree A0060406955 made by GENCI. This work has made use of the Horizon Cluster hosted by Institut d'Astrophysique de Paris; the authors thank St{\'e}phane Rouberol for running smoothly this cluster.

\appendix

\section{An estimate of the critical radius}
\label{sec:AnEstimateOfTheCriticalRadius}

In general, there exists no simple solution to Eq.~\eqref{eq:rcrit}. This still holds when the density profile is very simple such as a power-law ( Eq.~\eqref{eq:density}) for which Eq.~\eqref{eq:rcrit} reduces to Eq.~\eqref{eq:rcrit_polynomial}. However, in some situations ($\gamma=0;1;2$), Eq.~\eqref{eq:rcrit_polynomial} is a polynomial with simple solutions (which we do not report here) and $r_c$ can be expressed.

\begin{figure}
\includegraphics[width=\columnwidth]{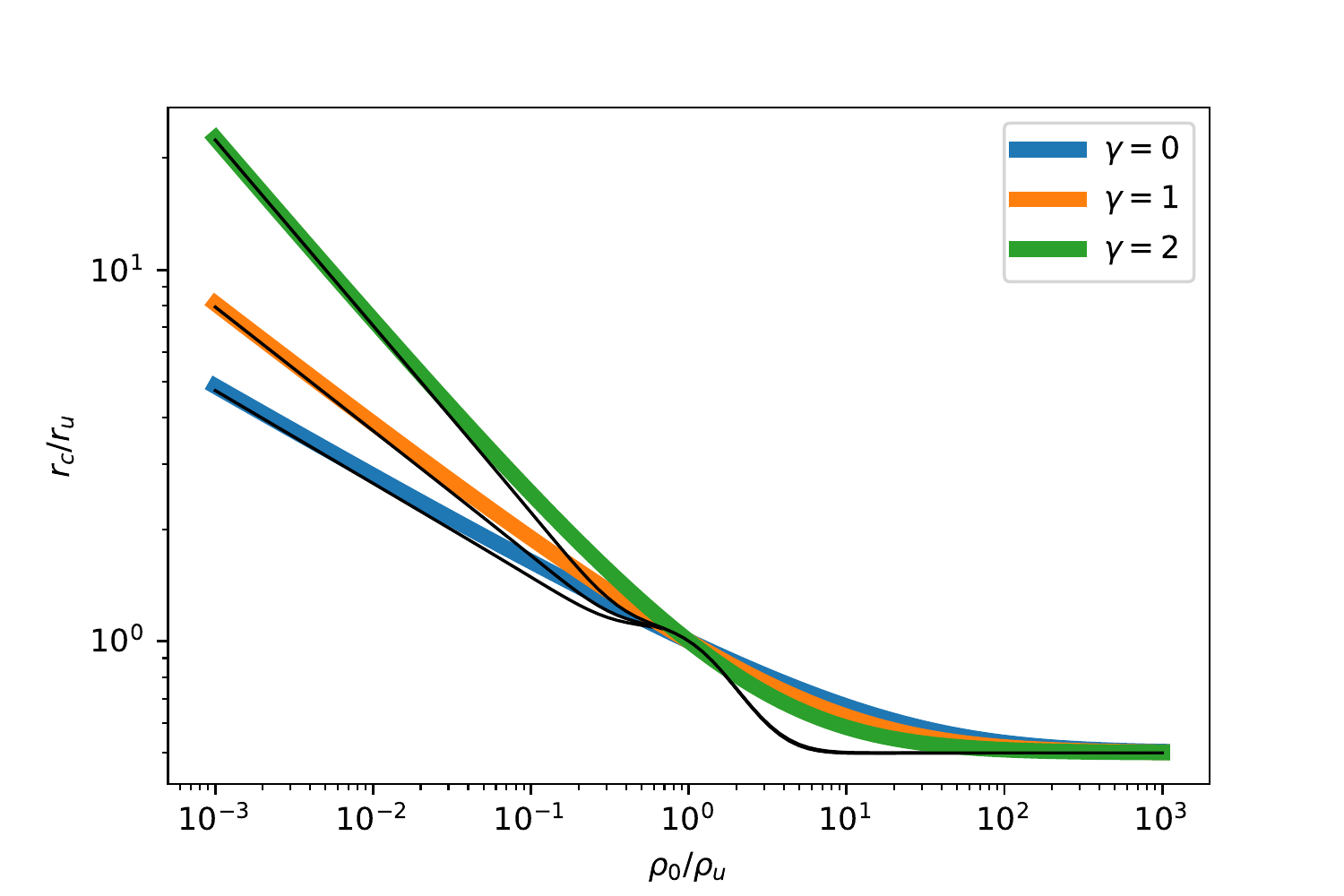}\\
\includegraphics[width=\columnwidth]{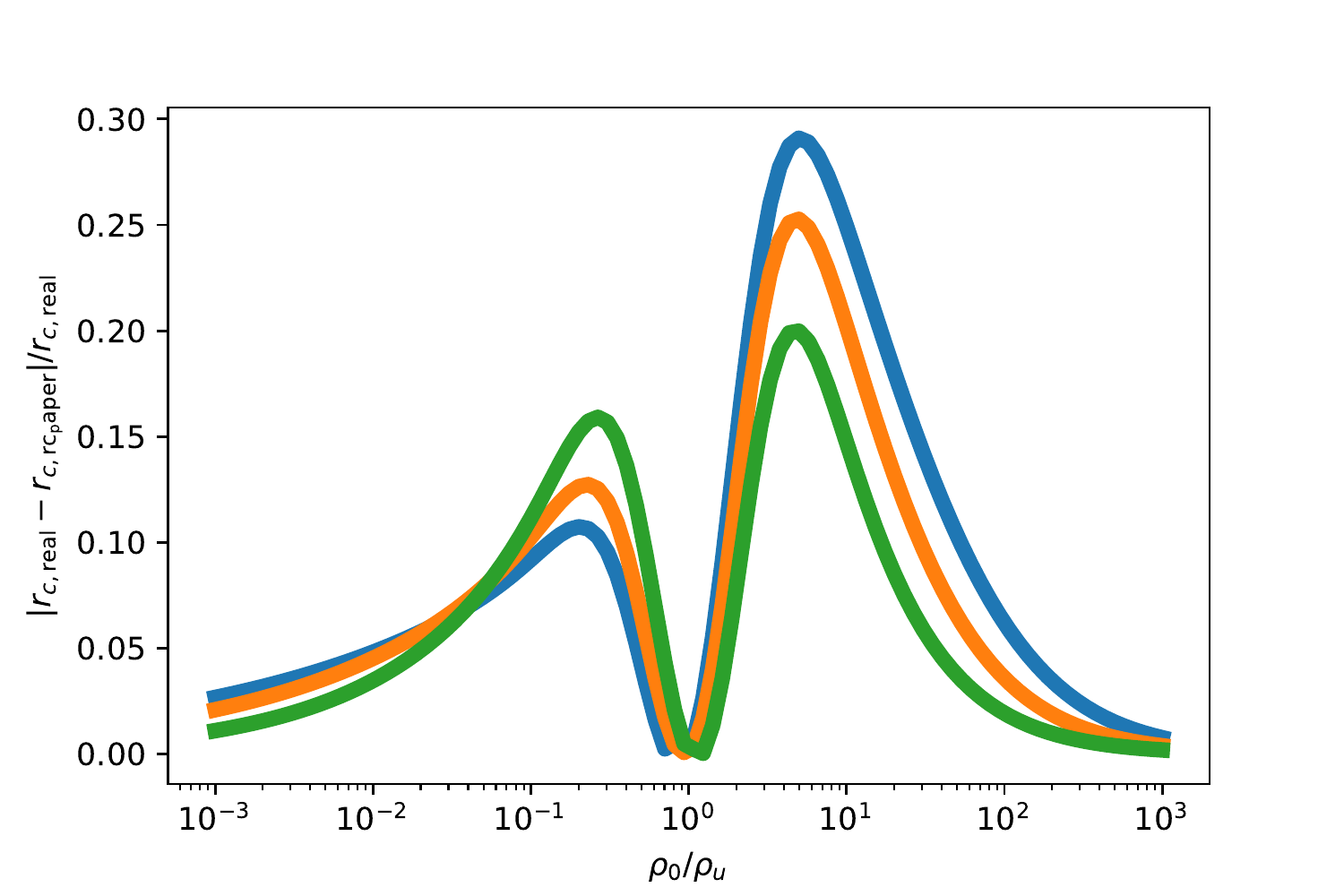}
\caption{\textbf{Top:} Exact solution of Eq.~\eqref{eq:rcrit_polynomial} (thick lines) and approximate solution from Eq.~\eqref{eq:rcrit_fit}. \textbf{Bottom:} Relative difference between $r_{c,{\rm real}}$, the real solution of Eq.~\eqref{eq:rcrit}, and $r_{c,{\rm fit}}$, the approximate value given by  Eq.~\eqref{eq:rcrit_fit}.}
\label{fig:rcApprox}
\end{figure}

In Fig.~\ref{fig:rcApprox}, we show the in the top panel the exact solution solving the polynomial (thick lines) and our approximate solution given by Eq.~\eqref{eq:rcrit_fit} (thin lines), and in the bottom panel relative difference between solutions. For $\gamma$ spanning between 0 and 2, \ie almost all the value allowed in our subgrid model, and for $\rho_0  / \rho_{u}$ spanning 6  orders of magnitude, the relative diffence peaks  at  30\%, which we consider as ``reasonable'' given the assumptions of the model.

\section{Effective radius of galaxies}
\label{sec:EffectiveRadiusOfGalaxies}

Contrary to halos, for which the virial radius can be defined to obtain the ``size'' of the structure, there are no clear definition for the size of a galaxy. In this Appendix we define the effective radius $R_\mathrm{eff}$ which we use for the ``size'' of the galaxy.

Once gravitationnally bound structures have been detected with \textsc{AdaptaHOP}, we compute the pseudo--inertia tensor:
\begin{eqnarray}
\widetilde{I}_{ij} = \sum_k m_k x_{i,k} x_{j,k} \, ,
\end{eqnarray}
where the sum is made on stellar particles $k$ belonging to the galaxy, with masses $m_k$ and positions $(x_{1,k},x_{2,k},x_{3,k})=(x_k,y_k,z_k)$ from the center of the galaxy. 

From $\widetilde{I}$ we can obtain the principal ellipsoid of the galaxy. The eigenvectors are the principal directions, and the eigenvalues $(I_1,I_2,I_3)$ are related to the principal axis $(a_1, a_2, a_3)$ by:
\begin{eqnarray}
I_i = \frac{1}{5}M a^2_i \, ,
\end{eqnarray}
where $M$ is the total mass of the galaxy, and the 1/5 factor is added so that the equation is correct for a homogeneous ellipsoid.

Once the principal ellipsoid is known, we compute the mass in concentric ellipsoid and find the one which contains 90\% of the total mass of the galaxy. The principal axis of this ellipsoid are $(a_\mathrm{1, eff}, a_\mathrm{2, eff}, a_\mathrm{3, eff})=(\alpha_\mathrm{eff} a_1, \alpha_\mathrm{eff} a_2, \alpha_\mathrm{eff} a_3)$, $\alpha_\mathrm{eff} > 0$, so that the effective radius is given by:
\begin{eqnarray}
R_\mathrm{eff} = (a_\mathrm{1, eff} a_\mathrm{2, eff} a_\mathrm{3, eff})^{1/3} \, .
\end{eqnarray}

\bsp	\label{lastpage}
\end{document}